\documentclass[acmsmall]{acmart}

\usepackage{multirow}
\usepackage{enumitem}
\usepackage{makecell}
\usepackage{csquotes}
\usepackage{graphicx}
\newcommand{\emoji}[1]{\raisebox{-0.2ex}{\includegraphics[height=1.8ex]{emojis/#1.png}}}

\newcommand{\Sys}{\textsc{Creact}~}

\newcommand{\revised}[1]{#1}
\newcommand{\review}[1]{}
\newcommand{\will}[1]{#1}

\AtBeginDocument{%
  \providecommand\BibTeX{{%
    \normalfont B\kern-0.5em{\scshape i\kern-0.25em b}\kern-0.8em\TeX}}}

\setcopyright{cc}
\setcctype{by}
\acmJournal{PACMHCI}
\acmYear{2027} \acmVolume{11} \acmNumber{1} \acmArticle{GROUP48}
\acmMonth{1} \acmDOI{10.1145/3859434}

\begin{document}

\title[The Impact of Personalized AI Response Suggestion \\ on Content Creators' Direct Messaging]{The Impact of Personalized AI Response Suggestions on Content Creators' Direct Messaging}

\author{Eun Jeong Kang}
\email{ek646@cornell.edu}
\orcid{0000-0003-1347-7556}
\affiliation{%
  \institution{Cornell University}
  \country{USA}
}

\author{Jingruo Chen}
\email{jc3564@cornell.edu}
\orcid{0009-0007-1606-5780}
\affiliation{%
  \institution{Cornell University}
  \country{USA}
}

\author{Jadon Geathers}
\email{jag569@cornell.edu}
\orcid{0009-0002-2851-2457}
\affiliation{%
  \institution{Cornell University}
  \country{USA}
}

\author{Susan R. Fussell}
\email{sfussell@cornell.edu}
\orcid{0000-0001-8980-5232}
\affiliation{%
  \institution{Cornell University}
  \country{USA}
}
\renewcommand{\shortauthors}{Kang et al.}

\begin{abstract}
  Direct messaging is often employed by social media content creators to directly interact with followers. Those creators attempt to engage with followers in private communication, but they often encounter a large volume of messages. Through formative interviews with 15 creators, we found that creators frequently experience fatigue from writing suitable content, as they moderate audience expectations while concurrently integrating their media personas into responses. To explore how AI can address these challenges, we designed a system providing personalized, AI-generated response suggestions tailored to creators’ writing styles to maintain their media personas. In a comparison study with 13 creators that examines how the system affects response behaviors, creators valued the system’s ability to soften potential impoliteness and anticipated higher response rates, helping them maintain a professional image. However, creators reported additional labor in verifying the authenticity of AI outputs to mitigate risks caused by AI misalignment. Our results indicate that while AI suggestions enhance efficiency, they can undermine a creator's engagement in communication by shifting their attention toward monitoring AI. 
\end{abstract}




\received{31 March 2026}
\received[revised]{17 July 2026}
\received[accepted]{17 September 2026}

\maketitle

\section{Introduction} 

Direct messaging (DM), which refers to one-to-one private instant messaging activities on social media, has become one of the basic communication channels where social media creators (SMCs), or influencers, can interact with fans, clients, and followers over social media.
\revised{In addition to public reactions such as comments and likes, it} allows more personalized, private, and targeted one-to-one interactions for creators to strengthen intimate social bonds with audiences more directly~\cite{nyt_2021, vardhan2023dm}. 
\revised{In blurring the line between private and public communication, direct messaging is becoming an important channel where creators and followers can build parasocial relationships. At the same time, a large volume of interactions directly sent from multiple audiences can place a burden on creators to perform emotional, relational labor~\cite{professional, tilton2025intimate, kang2025understanding}.}

To increase the efficiency of message reactions, AI-based message writing tools have been proposed and adopted for social media platforms in the real world. For instance, response candidates such as "Thank you" are recommended to users so they can quickly respond to message senders~\cite{kannan2016smart}. Beyond response generation, automated agents act on behalf of creators to interact with message senders, answering frequent questions in their place~\cite{knibbsonlyfans}.
However, despite the growing use of AI-based tools to facilitate interpersonal communication, \revised{how these interventions can be designed to effectively assist creators in their work on social media remains underexplored.
Direct messaging requires interlocutors to be private and personal, while the interactions SMCs present toward followers are designed to convey authenticity and to balance personal and public boundaries~\cite{bonifacio2025self, duffy2017platform}.
By refining their self-presentation for target audiences—strategically and deliberately choosing visuals or text that articulate who they are—they aim not only to strengthen their authenticity but also to enhance their brand value, fostering intimacy and aspiration among followers~\cite{simpson2023rethinking, duffy2017platform}.
At the same time, they should protect themselves from unwanted interactions~\cite{thomas2022s}. AI interventions for message communication that are not designed around creators' AI usage behaviors may undermine their efforts to manage parasocial expectations and safely regulate creator–follower relationships.}

Through formative interviews with 15 creators, we found that their unique context induces specific response composition behaviors. These behaviors include integrating their media personas into messages and moderating their reactions to avoid over-engaging with message senders. This often resulted in overuse of time for responses and disengagement from DMs through short replies or non-responsiveness.
 This highlights that boundary setting -- negotiating tensions between personal and public personas in interactions~\cite{hamilton2023nudes} -- is raised in response composition and varies depending on creators’ channel themes and the sender's messages.
Additionally, creators commonly receive six types of direct messages, ranging from personal to business-related, which makes it challenging to adjust content appropriately and maintain clear persona boundaries.

Guided by the findings from formative studies, we designed \Sys, a system that aids creators in crafting DM responses. The system generates two personalized AI-based response suggestions aligned with creators' media personas, derived from their social media channel data and characterized by typical writing styles (e.g., the use of emojis, expressiveness, and formality levels). 
The system generates message suggestions or revises messages using Large Language Models (LLMs), informed by the creator's writing style, conversation history, and the sender's previous messages, to reduce the creator's burden in direct message exchanges.
These features enable creators to maintain consistency and customize their social media personas based on differing message types. This customization facilitates channel management and helps secure personal boundaries.

 In a comparison study involving 13 creators, we examined how this system reshapes creators' response behaviors and values, and the challenges of the system compared to their current practices of using a standard messaging interface alongside general-purpose AI tools.
Participants composing responses with \Sys produced longer messages and actively utilized AI features. Participants felt social self-confidence in adjusting their tones in writing responses, valuing the system’s ability to soften potential impoliteness and anticipating higher response rates. However, they also reported additional labor in verifying AI outputs and noted that persona-based customization could limit the expression of their individuality, potentially undermining relationship building with audiences. Overall, our research underscores the trade-offs involved in designing personalized AI response suggestions for writing direct messages for creators, highlighting the importance of accounting for creator–audience relationships and creators’ professional growth as long-term design goals.

Our contributions are as follows:
\begin{itemize}
    \item We present insights gained from SMCs' lived experiences regarding SMCs’ strategic self-presentation practices and the challenges they face in crafting DM responses.
    \item We propose a framework for a creator-facing direct messaging interface, \Sys, which offers personalized AI-based response suggestions to help SMCs compose messages aligned with their writing styles.
    \item We derive design insights for personalized AI messaging systems based on a comparative study evaluating creators' holistic experience.
\end{itemize}



\section{Related Work}
In this section, we introduce social media content creators' backgrounds and direct messaging communication channels, and how technologies, especially AI, mediate communication. 

\subsection{AI-Supported Message Writing in AI-mediated Communication}

 Building on traditional Computer-Mediated Communication (CMC), AI-mediated communication (AI-MC) refers to interpersonal interactions modified, augmented, or generated by AI to achieve specific goals \cite{hancock2020ai}. Researchers have extensively explored AI's role across diverse text-based contexts, including group discussions \cite{lee2020solutionchat}, the workplace \cite{miura2025understanding, kadoma2024role}, romantic relationships \cite{kim2019love, liu2021significant}, \revised{consent exchange~\cite{furlo2021rethinking}}, creative collaboration \cite{suh2021ai}, and cross-cultural exchange \cite{yong2025poet}. \revised{The existing research} primarily focuses on interpersonal communication among general users who interact to achieve shared \revised{communication} goals.

 Within these settings, AI manifests in distinct forms. Some interventions utilize explicit AI agents visible to all interlocutors to help ground conversation topics \cite{yong2025poet, suh2021ai}. Conversely, others provide implicit support through AI-generated reply suggestions in the back channel \cite{lee2020solutionchat}. For example, Swain et al. introduced Care-Pilot, a system that guides client-service representatives in composing empathetic text messages by analyzing the conversation context \revised{in the organizational context} \cite{das2025ai}. 
 Lee et al. \cite{lee2025amplifying} identified four \revised{AI-MC design patterns emerging in various communication settings} between interlocutors and AI assistance. Collectively, this body of work highlights the utility of AI in mitigating communication friction \cite{hohenstein2020ai} and facilitating collaboration by reducing psychological burdens \cite{suh2021ai}.

 However, \revised{the involvement of AI in communication settings} introduces significant \revised{relational} trade-offs \revised{for people who use AI and for how their messages are perceived by other communication partners~\cite{hohenstein2023artificial}}. 
 Previous research indicates that the disclosure of AI usage can undermine receivers' perceptions of message senders, reducing the perceived trustworthiness and authenticity of messages when AI is known to support the senders \cite{jakesch2019ai, glikson2023ai}. 
 Furthermore, the specific language style of AI suggestions -- such as their level of hesitancy \cite{kadoma2024role} or whether they operate at the sentence level \cite{fu2023comparing} -- can impact the sender’s sense of agency, ownership, and attitudes toward their writing. 
 In summary, this line of research highlights the importance of understanding how \revised{the design of AI supports} influences communication, impacting both the message sender and the receiver's perception.

 Scholarship suggests that AI-mediated support should be context-aware, incorporating social norms such as relational dynamics and conversation topics into its design \cite{robertson2021can, liu2022will}. 
 Early attempts at AI-supported communication, such as Google Allo, failed to account for these fluid conversation dynamics, leading to communicative breakdowns \cite{hohenstein2018ai}. Such failures often stem from the fact that interlocutors naturally perform audience design, adjusting their linguistic styles and code-switching to suit specific situational contexts \cite{fussell1989understanding, bell1984language}. 
 Building on this, Robertson et al. \cite{robertson2021can} emphasize that the appropriateness of AI suggestions is determined by the social context\revised{ that interlocutors share}; ignoring these norms can result in \revised{the sharing of misconceptions about AI-assisted message users, degrading} interpersonal relationships~\cite{mieczkowski2021ai}. Fu et al. \cite{fu2024text} demonstrate that AI support requirements fluctuate based on the formality and stakes of the interaction, which vary significantly across different user contexts. 
 With a consensus regarding the importance of these social and communicative contexts in designing AI-MC interventions, \revised{our research specifically focused on how AI might support users who use interpersonal communication as parasocial interaction in their occupational practices. Through design exploration of AI interventions built based on creators' needs in message composition, we examined the values and trade-offs that creators might have in their DM experiences.}

\subsection{Parasocial Interaction of Social Media Content Creators}

Social media platforms have become essential venues for professional social media content creators (SMCs) to maintain networks, promote their work, express themselves, and earn their livelihood~\cite{duffy2023platform, duffy2020algorithmic}. 
Within these platforms, user engagement -- such as likes and views -- serves as a critical signal for SMCs to receive feedback on their content and sustain their careers. 
Because platform algorithms prioritize highly engaged content for recommendation, and brand partnership is often decided based on their perceived audience groups~\cite{caplan2020tiered}, maintaining ties with fans and communities is unavoidable for maintaining their visibility~\cite{zysman2016next, choi2023creator, randle2015towards, simpson2023rethinking, bonifacio2025self, bishop2019managing}. 
Consequently, their strategic intimacy often cultivates parasocial interaction, which is an illusory, quasi-social bond where media consumers experience a perceived direct, bidirectional relationship between creators and followers~\cite{labrecque2014fostering, kang2025understanding}. For instance, creators may ``like'' follower comments or express direct appreciation within their content to foster these connections~\cite{rihl2019youtube}. By developing this psychological connection, followers experience a heightened sense of familiarity, closeness, and identification with the creator~\cite{kim2022keeping}.

SMCs leverage platform affordances to facilitate these parasocial interactions. The spectrum of using these affordances ranges from mass communication (e.g., broadcasting content) to dyadic interactions (e.g., replying to private messages), and often spans multiple platforms \cite{bonifacio2023beyond, arriagada2020you}. Scholars highlight that these interactions are often strategic and performative, designed to present creators as authentic, genuine, and attractive to followers \cite{randle2015towards, malhotra2024curate, simpson2023rethinking, hoose2024self}. As their branding strategies rely on self-presentation rather than a purely commercial persona, SMCs tailor authenticity by maintaining consistency in tone, content, and imagery to project a ``true self'' \cite{duffy2017platform, khamis2017self, arriagada2020you}. Furthermore, they signal ``ordinariness" and ``approachability" by expressing intimacy and self-disclosing behind-the-scenes lifestyles \cite{marwick2011tweet}.

However, these diverse strategies are often laborious and can be emotionally demanding for them, functioning as a form of ``relational labor'' \cite{baym2015connect, duffy2017platform}. 
 SMCs maintain ongoing connections with their followers as a reward of economical values of how they are endorsed by followers~\cite{baym2015connect}. SMCs often feel vulnerable to over-engagement from unknown audiences. 
 Consequently, they strategically manage the boundaries between their personas and personal lives by moderating follower interactions \cite{duffy2017platform, cunningham2019social, bonifacio2023beyond}. Nevertheless, these interactions can simultaneously evoke genuine, emotionally meaningful exchanges that serve as a source of motivation for the SMCs themselves \cite{bonifacio2023beyond}. 
 \revised{Hair noted that this relational work can reduce creators’ burden of producing revenue through their creative labor, while it may also commodify their lives and self-images~\cite{hair2021friends}.}
 To reduce the burden on creators managing parasocial interactions from their side, platforms have introduced AI-based features such as voice clone messaging interacting with fans \cite{kang2025can}, automatic reply completion~\cite{youtube_2020}, and AI-simulated agents~\cite{meta_2023} in various communication settings. 
 \revised{However, scholars have noted a potential tension between these platform-driven features and the nuanced cultivation of authentic creator-follower relationships~\cite{kang2025can}. These technically evolving environments can also intensify labor for SMCs as platform workers by requiring them to constantly update their communication styles and activities accordingly~\cite{arriagada2020you}.} Yet, there remains a lack of understanding regarding how interaction designs impact creator behaviors, especially in the context of one-to-one messaging where creators engage with each individual.

\subsection{Direct Messaging in Mass-personal Communication Channels}

 Direct messaging (DM) is a way of communicating one-to-one that facilitates instant, dyadic interaction within social media platforms \cite{newton2022instagram}. DMs combine the privacy and personalization of traditional instant messaging with the anonymity and reach of public platforms, allowing users to connect regardless of prior acquaintance \cite{french2017anybody, yang2025does}. 
This hybrid context enables users to frequently interact with various social groups, from close friend groups to strangers online~\cite{huh2023help}, supporting diverse use cases, ranging from peer emotional support and expert consultation to promotional activities \cite{ali2023getting, verbeij2024happiness, hong2021just}. 
In contrast to private interpersonal communication like email or private instant messaging \cite{xu2017emphasizing, 10.1145/3643834.3661625, 10.1145/3064663.3064679}, DMs occur within social media environments characterized by mass-personal communication, where users navigate the subtle boundaries between personal intimacy and mass exposure~\cite{o2018masspersonal}. 
This distinctive context exposes users to risks, while it increases the chances of accessibility of DMs. Users may suffer from scams, harassment, and unsolicited sexual content through targeted messages \cite{vardhan2023dm}. Vulnerable populations, such as youth, are particularly susceptible to personalized scams, necessitating safer interaction designs and technology-driven mitigation strategies based on linguistic and engagement cues \cite{huh2023help, razi2023sliding, ali2023getting}.

\revised{DMs are also used to reach out to SMCs—by fans or clients who want to foster ``legitimate friendships'' and coordinate partnerships with them \cite{goodwin2023political, bonifacio2023beyond, silalahi2021micro}. Because of the volume of DMs they receive, brand clients are coached to write messages strategically in order to get attention from the SMCs they want to contract \cite{milovsevic2022instagram}. 
SMCs also reach out to other SMCs to collaborate and look for emotional support with individual fans; at the same time, they are privately attacked by unknown individuals \cite{parent2026victoria, global2026its}.}
Previous research has focused on patronage-based platforms where tiered personal interactions are facilitated through paid messaging services \cite{hamilton2023nudes, bonifacio2023beyond, bonifacio2025self}. In this context, an overwhelming volume of incoming messages can hinder a creator’s ability to maintain authentic interactions \cite{bonifacio2025self}. Also, Hamilton et al. found that sexual content creators utilize platform controls, such as blocking subscribers, to moderate misconduct and manage relationship boundaries for their safety \cite{hamilton2023nudes}. 
In this process, creators unavoidably maintain tensions between monetization benefits and interaction burdens to retain popularity \cite{bonifacio2023beyond}.

\paragraph{Our Approach}
 While existing literature explores SMC messaging within patronage and sexual content contexts \cite{hamilton2023nudes, bonifacio2023beyond}, these contexts are often patronage-based and focused on sexual content creators, where communication practices are highly influenced by platform monetization policies and the sensitive nature of their content. There remains a limited understanding of AI’s role in supporting boundary-setting for a broader range of creator personas. To address this gap, we adopted a Research through Design (RtD) approach \cite{zimmerman2007research} consisting of three phases: 1) investigating current behaviors and pain points; 2) developing a persona-aware system; and 3) assessing the prototype to elicit design refinements. Throughout this process, a professional content creator served as a co-author, providing expert feedback to ensure the system remained deeply grounded in the content creator context.

\section{Formative Interview}
\label{sec:3.3}
\revised{Previous research implies that SMCs may selectively choose strategies in online interactions\cite{silalahi2021micro} and manage their relational boundaries by regulating their interactions with followers \cite{bonifacio2025self}. Building on this, we specifically investigated how these strategies are presented in their message composition for DMs given their occupational context and how their experiences can inform design knowledge. Therefore,} our research question was:
\begin{itemize}
    \item \textbf{\revised{RQ1: How do SMCs write and send direct messages to followers in practice, and what challenges do they encounter in this process when they receive more messages than they can reasonably manage?}}
\end{itemize}

We conducted 15 IRB-approved semi-structured interviews with SMCs to understand how SMCs engage in direct messaging and the challenges they experience in responding to message senders in practice. This study extends our prior work~\cite{kang2025understanding} on creators’ DM practices, in which we presented a descriptive analysis of the study.

\subsection{Study Setup}
\label{formative_procedure}

We recruited SMCs based in the United States who had over 10,000 followers and actively used direct messaging features on social platforms (e.g., Instagram, TikTok, and X). Recruitment was conducted via social media platforms (e.g., Reddit), crowd-sourcing platforms (Prolific), and cold emails (Appendix~\ref{appendix:1}). 
The \revised{recruitment criteria} were informed by two pilot interviews with creators, prior literature~\cite{choi2023creator, silalahi2021micro}, and the feasibility of recruiting eligible participants. 
\revised{Since SMCs that have a huge number of followers (over 100k followers as macro-celebrities) might rely on human agents or not use direct messaging services~\cite{silalahi2021micro}, we decided to set our minimum threshold as a micro-celebrity who can manage a moderate number of followers.} We discovered that users with fewer followers usually had less difficulty managing large volumes of direct messages. This threshold allowed us to focus on recruiting participants who are more likely to face significant communication challenges.

We adopted a semi-structured interview-based approach given the highly contextual nature of messaging practices, which are often shaped by a creator's channel theme, posting tone, and audience demographics.
\revised{For the first phase, drawing upon prior work that focused on handling massive content from audiences~\cite{jhaver2022designing, park2019opportunities}, we conceptualized direct messaging as two sub-behavioral actions: 1) interpreting the sender's messages and 2) writing responses. 
To gain an understanding of their context, we began by asking participants about the common messages they receive through their creator accounts and how they perceive message senders. Focusing on each sub-process, we asked about their decision-making process and any challenges encountered.} 
With participants' consent, they were encouraged to open their DM inboxes and provide examples from their message history. \revised{We asked them to read out the content of several actual messages they typically received.}

\subsection{Data Analysis}
Interviews were audio-recorded and transcribed via Otter.AI~\cite{otterai}. After anonymizing the transcripts, we conducted a thematic analysis~\cite{braun2006using} using MAXQDA~\cite{maxqda}.
Two authors independently reviewed the transcripts to gain a holistic understanding, followed by open coding~\cite{wicks2017coding} to capture patterns and variations. We then compared initial coding schemes, identifying areas of convergence and divergence while using memo-writing to reflect on emerging interpretations. 
Through multiple rounds of discussion, we iteratively refined and merged codes. Finally, guided by our research questions, we synthesized these codes into higher-level themes capturing creators’ practices and perceptions. 

\revised{To identify DM typology specific to content creators, we reviewed DM cases that participants shared or described during the studies. We focused on the types of DMs they received, descriptively analyzing the purposes of message senders based on the message content and participants’ descriptions.
To validate the typology initially derived from our interviews, we further examined DM cases publicly introduced online by SMCs, following the approach of Das et al.~\cite{das2025ai}. We collected 203 DM excerpts voluntarily disclosed by creators in public and formative contexts, drawn from 10 YouTube videos with titles containing “DM” or “Direct Message.”
Two researchers reviewed the typology descriptions and then categorized the collected cases. The initial inter-rater reliability was 0.905. We iteratively refined the descriptions through discussion.}

\subsection{Key Insights}

\subsubsection{\revised{A Broad Spectrum of Direct Messages, from Personal to Professional}}
\label{sec.3.2.1}

We identified six types of direct messages SMCs commonly receive, ranging from general social interactions to professional inquiries \revised{(Table~\ref{tab:dm_type}). \revised{Chat conversations between creators and followers are mostly initiated and sustained by followers, who ask questions or share their thoughts. Note that these categories are not mutually exclusive within a conversation. Followers often send multiple DMs in succession that span more than one category.}
The `Personal Inquiries' category involves content that expresses intimacy or personal requests. When creators and message senders are mutual friends or family, creators feel less uncomfortable. However, when such messages come from followers, they are more likely to cross personal boundaries that creators do not wish to share publicly. In contrast, `C}reator-persona-related Inquiries' and `Partnership Suggestion' \revised{are related to a creator’s professional identity and creative work. These types of messages include questions bound to a social media persona that creators construct on social media channels through their curated content. }
The cognitive and emotional demands of these messages vary significantly. While some are low-effort random reactions, others -- such as `Content Inquiries' and `Content Critiques' -- require further knowledge and professional judgment \revised{(e.g., asking a health news influencer how to cope with Covid 19).} 
Echoing prior work~\cite{bonifacio2023beyond}, participants reported that repetitive or negative content often causes significant emotional and cognitive distress, highlighting the burden of managing diverse audience interactions.

\subsubsection{Balancing Reaction Management Across Different Message Types based on Channel Goals}
\label{sec.3.2.2}

Upon receiving a DM, participants rapidly assess the content and sender to prioritize responses based on their channel goals. They favor messages relevant to their specific themes, such as travel tips for a cultural creator, and minimize latency for mutual followers or potential brand clients to secure financial or relational benefits. This selective engagement allows creators to strategically manage limited resources while maintaining an audience that supports their objectives.
P3 mentions, \textit{`` I'd say the challenge is sometimes it's (direct messaging) tiresome, maybe you can get 100 messages, and then they expect you to respond to them. It's also very streaming. Because someone sent me a very good message, acknowledging me, maybe giving me some good advice. But I'm not able to respond to those messages because maybe I never even saw them. ''}
Participants did not consider all messages worthy of a response; irrelevant or uncomfortable content was typically ignored or de-prioritized. P1, for instance, found messages requesting intimate personal details—such as a phone number—inappropriate. To prevent further conversation, he often ignored these DMs or provided brief, non-committal responses.

\begin{table*}[ht]
\centering
\caption{Direct message typology and examples SMCs typically received from audiences, drawn from interviews}
\label{tab:dm_type}
\resizebox{0.98\textwidth}{!}{%
\begin{tabular}{m{3cm}m{7cm}m{5.5cm}}
\hline
\textbf{DM type} & \textbf{Description} & \textbf{Examples}  \\ \hline

Personal Inquiry &
\begin{itemize}[noitemsep, topsep=5pt, leftmargin=1em]
\item Efforts to establish friendships beyond online interactions (i.e., seeking personal information) or request personal favors to utilize creators' popularity
\end{itemize}
&
\begin{itemize}[noitemsep, topsep=5pt, leftmargin=1em]
\item ``Can I see more of you?''
\item ``I think short hair suits you.''
\item ``Can you give me contact information?''
\end{itemize} \\ \hline

Creator Persona-related Inquiry &
\begin{itemize}[noitemsep, topsep=5pt, leftmargin=1em]
\item Efforts to obtain detailed information on the lifestyle and workflow of creators based on their social media posts \textit{(Channel-related questions)}
\item Seek advice of message senders, expecting to share a point of view that reflects creators' lifestyle \textit{(Creator lifestyle-related questions)}
\end{itemize}
&
\begin{itemize}[noitemsep, topsep=5pt, leftmargin=1em]
\item ``What product do you use to give your lips a glowing appearance?''
\item ``Would you like to recommend brands?''
\item ``Could you provide the recipe for the lemon pie posted yesterday?''
\end{itemize} \\ \hline

Partnership Suggestion &
\begin{itemize}[noitemsep, topsep=5pt, leftmargin=1em]
\item Suggestion for collaborations in which creators can get benefits from collaborators (e.g., clients, creators) \textit{(Benefit-sharing collaborations)}
\item Proposals for asking for free labor for promotion to utilize creators' popularity \textit{(Unpaid promotional partnerships)}
\end{itemize}
&
\begin{itemize}[noitemsep, topsep=5pt, leftmargin=1em]
\item ``We'll send you a product.''
\item ``We love your content, and we would like to send you this.''
\end{itemize} \\ \hline

Compliment &
\begin{itemize}[noitemsep, topsep=5pt, leftmargin=1em]
\item Positive comment about creators' social media posting content, lifestyles, and editing styles, showing how message senders appreciate creators' activities
\end{itemize}
&
\begin{itemize}[noitemsep, topsep=5pt, leftmargin=1em]
\item ``Yo man! You're super hilarious and your videos never failed to make me laugh. Keep it up''
\end{itemize} \\ \hline

Content Critique &
\begin{itemize}[noitemsep, topsep=5pt, leftmargin=1em]
\item Suggestions that creators enhance their channels by providing feedback on their posting and editing styles, as well as their choice of attire.
\end{itemize}
&
\begin{itemize}[noitemsep, topsep=5pt, leftmargin=1em]
\item ``Shoutout, would you and Lin ever make a podcast together?''
\end{itemize} \\ \hline

Random &
\begin{itemize}[noitemsep, topsep=5pt, leftmargin=1em]
\item Sending content (images, anecdotes) the creator does not agree is relevant to them
\end{itemize}
&
\begin{itemize}[noitemsep, topsep=5pt, leftmargin=1em]
\item Sending videos of themselves playing different violins to music creators
\end{itemize} \\ \hline

\end{tabular}%
}
\end{table*}

\subsubsection{Stylizing Responses Tailored for Imagined Audience \revised{Groups} and Channel Personas}
\label{sec.3.2.3}
Participants consistently presented specific personas through their responses (e.g., writing styles), echoing prior findings~\cite{simpson2023rethinking}.
By default, most participants maintained professional and intimate personas when writing responses. Notably, P10 and P15 adopted different strategies, aiming to appear more straightforward in their messages; they believed this approach could help shield them from harassment by strangers. 

These personas were not applied uniformly across all messages.
Participants sometimes adjusted message tone and response detail based on how they anticipated message senders would react. For example, when responding to clients with whom they wished to continue conversations, participants maintained a formal tone and occasionally relied on prepared content (e.g., contact information) to avoid mistakes. 

Additionally, rather than strictly adhering to fixed persona characteristics, participants prioritized tailoring their responses to different audience groups, aligning message tone and content with the traits they considered most appropriate for their prioritized audiences. \revised{When they assumed a message sender was a `brand client,' based on their message content and authenticated channel, they maintained formal tones, trying to present their professional experience. P9 stated he sometimes used ChatGPT to polish his tone in a message response. }P5 (Fashion) stated,\textit{ ``With my followers, I stick with my TikTok persona. With clients, I'm very professional and straight to the point. Then with my friends and family, I'm just casual, whatever. There's like, three different ways of how I talk to each person''}

In this process, participants experienced difficulties in choosing appropriate wording and tone that conveyed their intent without offending or upsetting specific audience groups, while maintaining personas that appeared authentic and approachable. This challenge was particularly salient in situations requiring successful conversational grounding, such as declining offers, negotiating deals, or consulting with message senders. These findings indicate that creators struggle with audience design in their responses due to insufficient information about their audiences~\cite{fussell1989understanding}.
P1 and P9 expressed a desire to use language that makes them appear open and approachable. However, they found it challenging to decline certain requests, such as sharing personal opinions, while appearing authentic.

\subsubsection{Self-Control Reactions to Avoid Over-Engagement of Individuals}
\label{sec.3.2.4}

Participants believed that the information shared should vary based on their familiarity and priority for audience groups. They explicitly mentioned `personal boundary', echoing existing research illustrating the unseen boundaries and limits they established with some audience groups~\cite{bonifacio2023beyond, hamilton2023nudes}. When they write messages, they calculate the potential interaction effort they should further spend on the message receiver. 
P14 mentioned that it was challenging to restrict the message receiver from asking for personal information while maintaining a friendly atmosphere in their messages. He said, \textit{``I really want to be relatable to my general audience and everyone following me, but the internet is a really weird place, and people start feeling as if they know you that much.''}

\begin{table*}[ht]
  \centering
  \caption{\revised{Mapping between formative study insights, design goals, and \Sys features}}
  \label{tab:mapping}
  \resizebox{0.98\textwidth}{!}{%
  \begin{tabular}{m{4.2cm}m{4.2cm}m{2.6cm}m{5cm}}
  \hline
  \textbf{Key Insight} & \textbf{Identified Challenges} & \textbf{Design Goal} & \textbf{Design Components} \\ \hline
  \revised{Creators selectively prioritize reactions based on channel goals (\S\ref{sec.3.2.2})} &
  \revised{Limited time to identify prioritized messages; uncomfortable messages are ignored} &
  \revised{DG2, DG4} &
  \revised{Automatic DM classification into six categories; category-filtered inbox and thread summaries (\S5.3, \S5.3.1)} \\ \hline

   \revised{Creators stylize responses for imagined audience groups and channel personas (\S\ref{sec.3.2.3})}&
  \revised{Difficulty choosing wording and tone according to each audience group, staying in their social media persona without offending them} &
  \revised{DG1, DG3} &
  \revised{Persona construction from social media posts; editable persona panel; category-specific writing style settings (\S5.2, \S5.3)} \\ \hline

   \revised{Creators self-limit responses to avoid over-engagement and protect boundaries (\S\ref{sec.3.2.4})}&
  \revised{Detailed replies invite unwanted follow-ups; adequate-but-guarded responses are effortful} &
  \revised{DG3, DG4} &
  \revised{Enhance Message feature; Response Suggestion feature; per-category exclusion of
  personal data; review panel before sending (\S5.3.1, \S5.4)} \\ \hline

  \end{tabular}%
  }
  \end{table*}

Participants reported that highly detailed or specific responses could inadvertently invite unwanted follow-up interactions or harassment. Because crafting lengthy, empathetic messages is exhausting and often encourages further engagement from strangers, creators found it challenging to respond adequately without disclosing specific information. Consequently, to disengage from unnecessary messages and limit conversation length, participants like P1, P9, and P6 intentionally kept responses brief or relied on emojis to avoid prompting additional questions.


\section{Design Goals}


Drawing on insights about their behaviors obtained from formative studies \revised{(Findings \ref{sec.3.2.2} - \ref{sec.3.2.4})} and literature work, we formulated four design goals for direct messaging tools tailored to creators for response composition. \revised{In particular, we specifically look into how each key insight can inform design components supporting writing response composition in DM.}


\textbf{DG1.} The system should align with the creator’s curated social media persona by reflecting their specific text, visual, and interactional cues \revised{on their channels}. This ensures a coherent professional image across all communication contexts (Section~\ref{sec.3.2.3}). 

\textbf{DG2.} The system should allow creators to tailor AI assistance based on DM type and priority. This supports strategic reaction management according to varying situational demands and channel goals (Section~\ref{sec.3.2.2}). 

\textbf{DG3.} The system should support creators in composing responses that align with their intentions, particularly in situations where they struggle to employ appropriate language to convey their intentions \revised{(Sections~\ref{sec.3.2.3} and~\ref{sec.3.2.4})}.

\textbf{DG4.} To minimize response latency without sacrificing credibility, the system should enable creators to easily generate, review, and verify AI-generated outputs in order to reduce turn-taking latency. Such assistance should be grounded in the conversation context and require minimal effort from creators, allowing creators to correct errors \revised{(Sections~\ref{sec.3.2.2} and~\ref{sec.3.2.4})}.

\section{\Sys}
Based on the design goals, we designed \Sys, an AI-powered \revised{DM response writing} tool that features personalized AI-based response suggestions constructed with a creator's social media persona that comprises their social media channel characteristics and writing style preferences.
In this section, we provide an overview of the user interface of \Sys.

\subsection{Overview}

\begin{figure}
    \centering
    \includegraphics[width=1\linewidth]{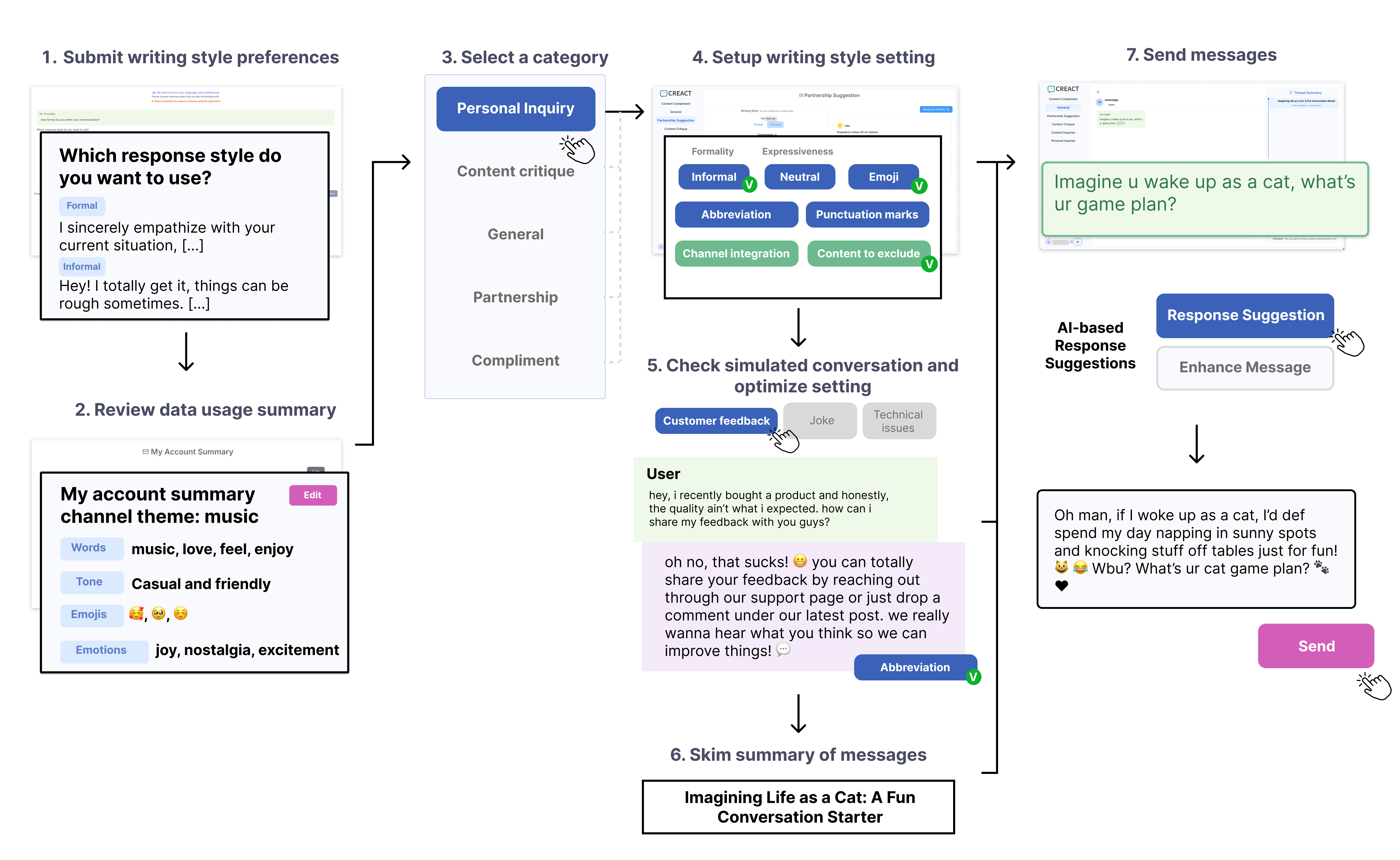}
    \caption{Overall user flow in the \Sys interface}
    \label{fig:interface}
\end{figure}

When first accessing the interface (Figure \ref{fig:interface}), creators input their writing style preferences (1). A left-side panel then provides access to categorized messages and a social media persona summary (2). \revised{On the social media persona summary page, creators can see textual social media data and edit} AI-extracted attributes used for personalization.
Selecting a category (3) reveals two sub-sections: one for adjusting the desired writing style (4) with a preview via scenario simulation (5), and another listing category-specific message threads. After reviewing thread summaries (6), creators select and read individual messages (7).

In the message board \revised{(Figure \ref{fig:main_page})}, where creators mainly write responses for the message sender to assist with drafting, creators can use the `` Response suggestion" and ``Enhance message" features (Figure \ref{fig:main_page}-c,d). For instance, if creators draft a blunt response like "I do not want to accept the offer," they can use the ``Enhance'' tool to request a polite, simple alternative. The AI might then suggest: \textit{``I hope you understand, but I would prefer not to accept the offer right now."} After final \revised{manual edits by creators}, such as removing emojis, creators send the message.

\subsection{Constructing a Social Media Persona based on Social Media Content}

Our system utilizes a \textit{social media persona}, defined as the strategic impression creators project through selectively shared content \cite{leary2011self, enke2021social, miguel2024self}, to guide AI-mediated interactions. 
Building on research that extracts attributes from unstructured data to design persona-based LLM agents \cite{choi2025proxona, shin2025postermate}, our system leverages creators’ social media posts as informative data to identify creators' pervasive stylistic features (DG1; Figure \ref{fig:interface}-(2)). 
\revised{Technically, the system prompts an LLM to extract the ten most frequently recurring stylistic features from the creator's posts — spanning emotional register, tone, distinctive word choice, and emoji use — and returns them as a structured JSON object. An embedding model over the same textual data supplies additional context for response generation.}

\begin{figure}
    \centering
    \includegraphics[width=1\linewidth]{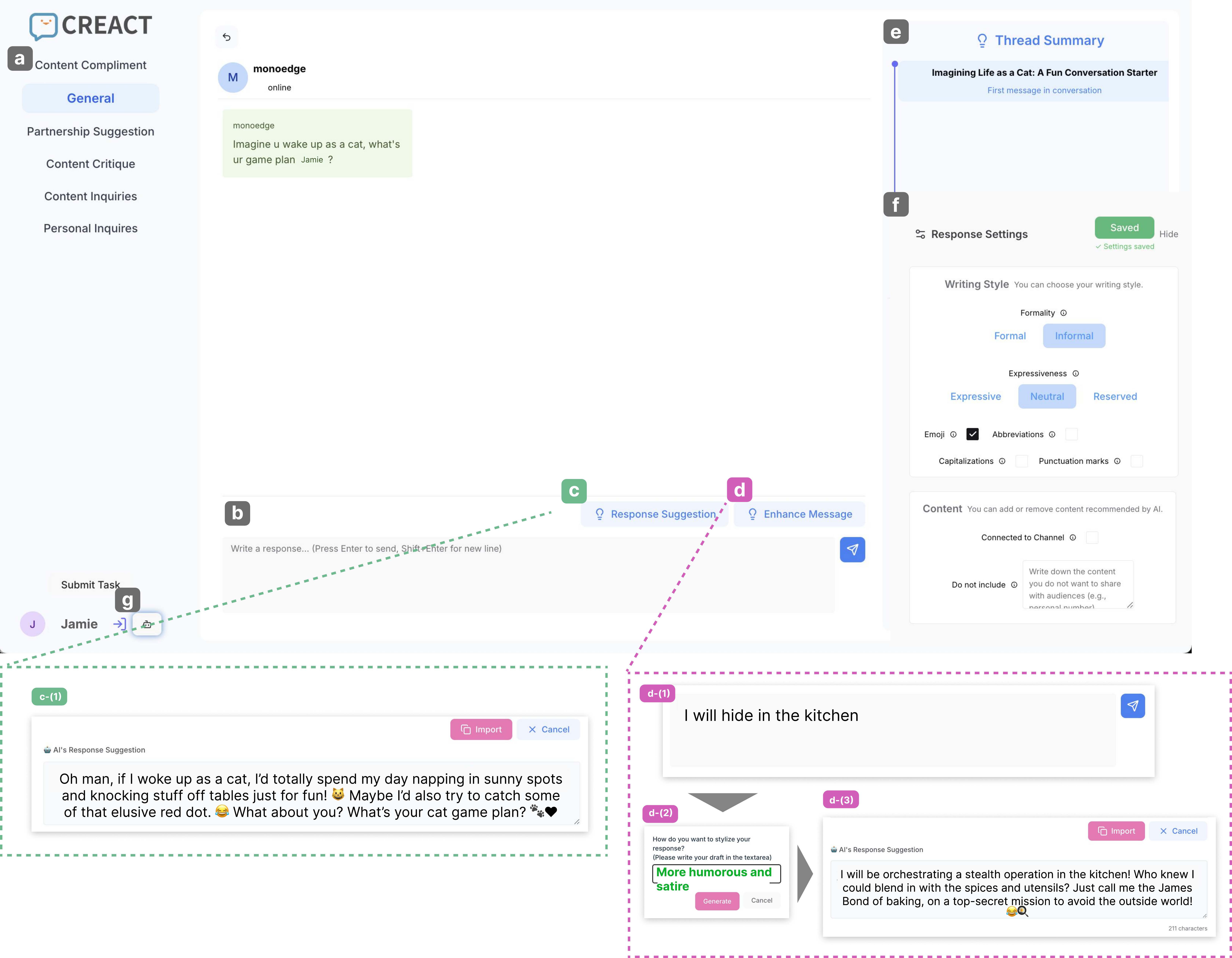}
    \caption{The snapshot of the chat conversation board where SMCs can send DMs in \Sys}
    \label{fig:main_page}
\end{figure}

To ensure transparency and controllability, the interface allows creators to verify or modify this extracted data via a side panel. By clicking their account profile Figure \ref{fig:main_page}-(g), creators can review, edit, or remove specific attributes, maintaining full agency over their digital persona and ensuring that data sharing remains voluntary and adjustable.

\subsection{Adopting Persona Styles and Content to Use for Varying DM Types}

Using the DM typology described in Table~\ref{tab:dm_type}, the system automatically classifies incoming DMs into six categories, enabling creators to prioritize responses and apply category-specific settings (DG2). Selecting a category in the interface filters the inbox and opens the Writing Style Settings and Scenario Simulation panels.

\revised{We classify by DM type rather than by inferred audience group. Because creators' boundaries around personal and professional disclosure are tied to who is asking, audience-group classification could in principle support finer-grained, audience-specific boundaries. We opted against it for two reasons. First, a fixed category set makes misclassifications easier for creators to notice and correct, since the categories are legible and stable, where an unnoticed false-positive case could surface information the creator intended to withhold. 
Second, inferring audience groups would entail collecting sender information beyond the message itself, raising privacy concerns.}

\subsubsection{Writing style setting}

Creators use Writing Style Settings to customize how their persona is reflected in AI-generated responses across different DM categories. We define "writing style" as a combination of linguistic cues, tone preferences, and content grounded in the creator’s prior social media posts. Incorporating these elements enhances the individuality of the text, reflecting the creator’s authentic self-presentation \cite{jensen2003we, albawardi2025self}.
Based on factors known to influence text-based communication—expressivity, formality, and punctuation— \cite{kraut1990informal, buschek2018personal, chen2021emoji, 2025lowercase}, creators instantiate those settings when they access the interface for the first time (Figure \ref{fig:interface}-1). 
Then in each DM category, creators toggle specific style elements according to their preferences. For instance, creators select a level of Expressivity (Reserved, Neutral, or Expressive), while other elements, such as emoji use or capitalization, are controlled via boolean toggles (Figure \ref{fig:interface}-4).
They can determine whether the AI should ground responses in their actual social media content. 
\revised{This feature enables boundary-setting across DM types by allowing creators to intentionally adjust the amount of information and language styles. C}reators may exclude specific personal or professional data from certain DM categories to maintain privacy or professional distance. \revised{Since creators can have different writing styles within a DM type depending on relational closeness (friend, client, and fan), t}hese settings can also be adjusted within individual conversation logs to ensure the AI's output remains aligned with the specific context of the interaction (Figure \ref{fig:main_page}-f). 

\subsubsection{Scenario Simulation}
To ensure stylistic accuracy, creators can use Scenario Simulation to preview AI-generated turn-taking conversations (Figure ~\ref{fig:interface} - (5)). Using LLMs, the system generates relevant topics (e.g., collaboration proposals) and simulated chat logs based on the current persona settings. This allows creators to iteratively optimize their settings and verify that the AI’s tone aligns with their desired self-presentation before engaging with actual audiences.

\begin{figure}
    \centering
    \includegraphics[width=1\linewidth]{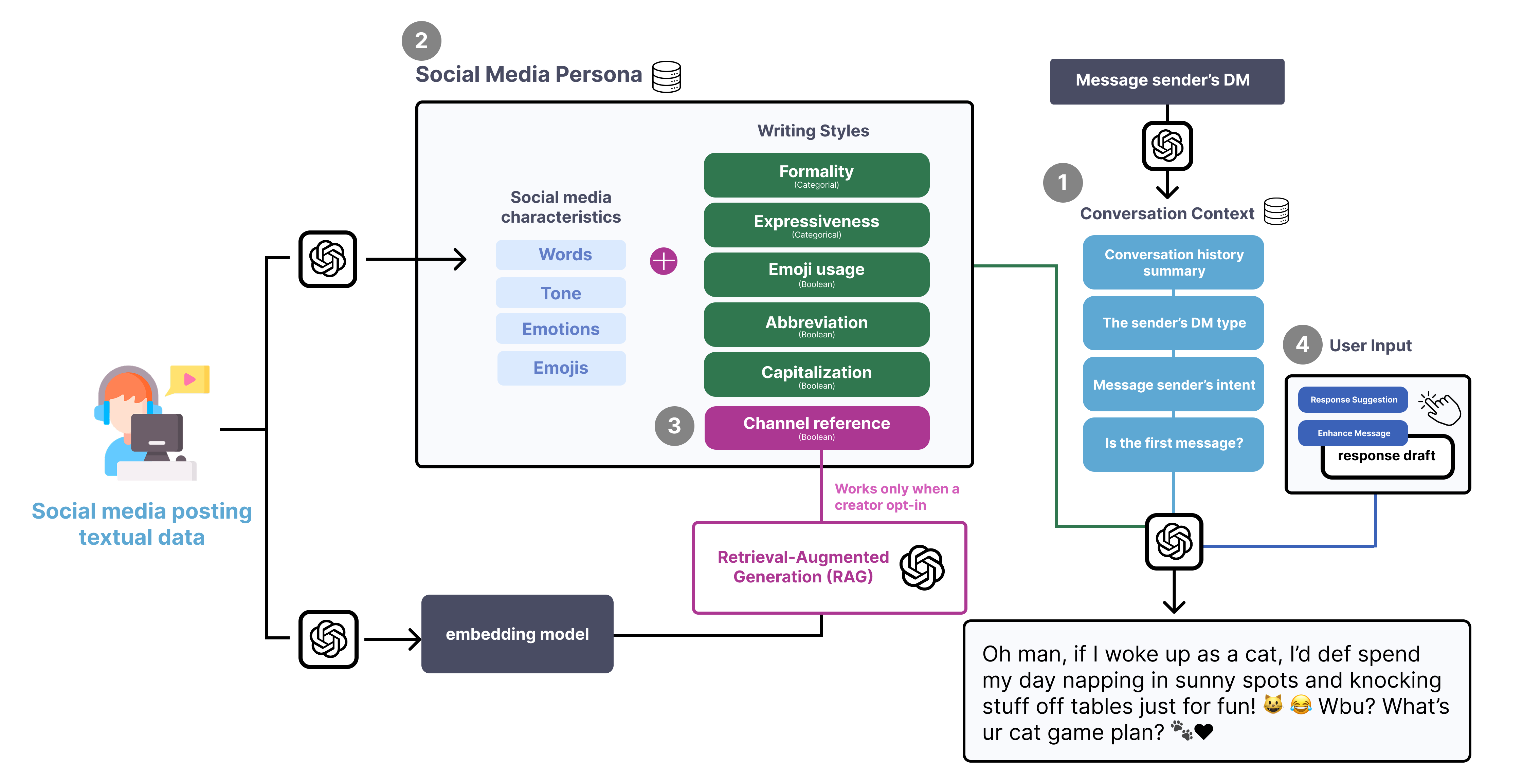}
    \caption{The pipeline to generate two personalized AI-generated response suggestions}
    \label{fig:pipeline}
\end{figure}

\subsection{Generating Personalized AI-based Response Suggestions}
\label{sec:5.4}

Sharing conversation context between interlocutors is important to ground effective communication~\cite{cho2020role, gergle2013using}. Also, chat-based interactions constitute relational communication activities that require message senders to continuously anticipate recipients’ knowledge, expectations, and interpretations~\cite{walther1992relational, hohenstein2018ai}. 

Guided by these insights and design goals (DG3 and DG4), we designed two LLM-based response suggestions that help creators when composing DM responses. Creators can interact with two features: \textit{Response Suggestion} and \textit{Enhance Message}. The \textit{Response Suggestion} generates a full candidate reply for efficiency. In contrast, the \textit{Enhance Message} allows creators to draft a message and use the AI to refine tone and phrasing. 
We designed these two complementary features to accommodate varying levels of creators’ experience in direct messaging, while maintaining authorial agency~\cite{fu2023comparing, fu2024text}. 

We developed a response generation pipeline that produces AI-based suggestions grounded in prior message turns and creators’ social media personas (Figure~\ref{fig:pipeline} -1, 2). Specifically, response suggestions are generated based on five elements: a summary of the ongoing conversation history; the DM type inferred from the previous message; the message sender’s communicative intent; whether the message initiates a new conversation; and a creator’s social media persona. These elements are incorporated into structured prompts that enable the system to generate responses appropriate to the interaction history while adapting to creators’ preferred self-presentation.

Upon receiving a DM, the system extracts conversational context (Figure~\ref{fig:pipeline}-1) using an LLM via prompt engineering and stores it in a database. When creators trigger a suggestion (Figure~\ref{fig:pipeline}-4), the LLM synthesizes this context with stored persona features (Figure~\ref{fig:pipeline}-2) to generate an AI-based response. When creators use \textit{Enhance Message}, their draft is passed to the LLM along with the extracted contextual elements to produce a revised response aligned with the ongoing interaction. A dedicated review panel enables creators to verify and revise all AI outputs before sending, ensuring credibility and human oversight.

When creators opt in to using channel references as part of their writing style preferences (Figure~\ref{fig:pipeline}-3), the system additionally employs a Retrieval-Augmented Generation (RAG) approach~\cite{lewis2020retrieval}. In this mode, an embedding model retrieves relevant snippets from the creator’s prior posts, which are then integrated into the generation prompt. This ensures that AI suggestions are not only stylistically consistent but also factually aligned with the creator’s established public discourse.

\subsection{Implementation}
The front-end was developed using the NextJS 16 framework and TypeScript, while the back-end utilized the Django REST framework, both served via Heroku. We employed OpenAI's GPT-4 model to implement AI-based response suggestion features through the LangChain framework (ChatModel and Embedding Model). The prompts for response suggestions and social media extractions are provided in Appendix~\ref{appendix:2}.
\begin{table*}[h!]
\centering    \caption{Demographic Information of Participants. All participants worked based on Instagram accounts}
\resizebox{0.7\textwidth}{!}{
    \begin{tabular}{lcccc}
        \toprule
        \multicolumn{1}{l}{\textbf{PID}} & \textbf{Channel Theme} & \textbf{Gender} & \textbf{\# of Followers} &   \textbf{Message Replying  Frequency} \\
        \midrule
        U1 & Fashion & M & 16.5k  & Once a week\\
        \hline
        U2 & Lifestyle & M & 19k  &   2-3 times a week\\
        \hline
        U3 & Lifestyle & M & 958k &  2-3 times a week\\
        \hline
        U4 & Comedy & M & 88.8k  &   more than 7 times a week\\
        \hline
        U5 & Antique & M & 11.1k  &  more than 7 times a week\\
        \hline
        U6 & Music & M & 13.1k & 4-6 times a week\\
        \hline
        U7 & Clothing & F & 114k  &  2-3 times a week\\
        \hline
        U8 & Lifestyle & F & 11.8k  &  more than 7 times a week\\
        \hline
        U9 & Couple & M & 324k  & more than 7 times a week\\
        \hline
        U10 & Clothing & F & 836k  & 4-6 times a week\\
        \hline
        U11 & Lifestyle & M & 32.3k & more than 7 times a week\\
        \hline
        U12 & Entrepreneur & F & 239k  & more than 7 times a week\\
        \hline
        U13 & Lifestyle & M & 10k &  more than 7 times a week\\
        \hline
    \end{tabular} 
    }\label{tab:participants_main}
\end{table*}

\section{Evaluation}
 We aimed to understand how our system reshapes SMCs' response composition behaviors, how SMCs use our system, and how it is valued in supporting their reaction management. 
We approached this with a mixed-method study to investigate the effectiveness of our approach, presenting \Sys to 13 SMCs who actively interact with their followers through their social media channels.

Our research questions are:
\begin{itemize}
    \item \textbf{RQ2.1: How do SMCs use an integrated, personalized AI messaging system?}
    \item \textbf{RQ2.2: \revised{What values and challenges do SMCs perceive in using personalized AI message suggestions in \Sys?}}
\end{itemize}

Participants were first asked to write message responses in a controlled condition and to provide qualitative feedback for our prototype based on their experience. In this section, we describe our study methods and report findings from this user study.

\subsection{Participants}
Similar to the criteria in the prior study, we recruited 13 SMCs (U1 - U13) who operated their own channels with more than 10,000 followers in the United States.
We invited participants who use Instagram as their main channel to communicate with followers, in addition to other channels. \revised{Since the prototype uses textual data that reflect creators’ channel characteristics and activities, we chose Instagram, where creators can introduce themselves through profile descriptions and posts. Although X users produce textual data substantially, we assumed that platform was likely to narrow our participant pool.}
One participant was involved in the earlier formative study, and others were all newly recruited. 
Table \ref{tab:participants_main} presents demographic information and Instagram channel information. 

All participants reported that they have used AI tools (i.e., ChatGPT, Smart Replies) when they wrote direct message responses (Median = $3.0$ on a 5-point ordinal scale -- 1 = Never, 5 = Always). They expressed strong willingness to adopt AI technologies for their creative work  (Median = $6.0$ on a 7-point Likert scale -- 1 = Strongly disagree, 7 = Strongly agree). Also, they expressed strong importance of designing how they appear to audiences on social media (Median = $6.31$ on a 7-point Likert scale -- 1 = Strongly disagree, 7 = Strongly agree)

\begin{figure}
    \centering
    \includegraphics[width=1\linewidth]{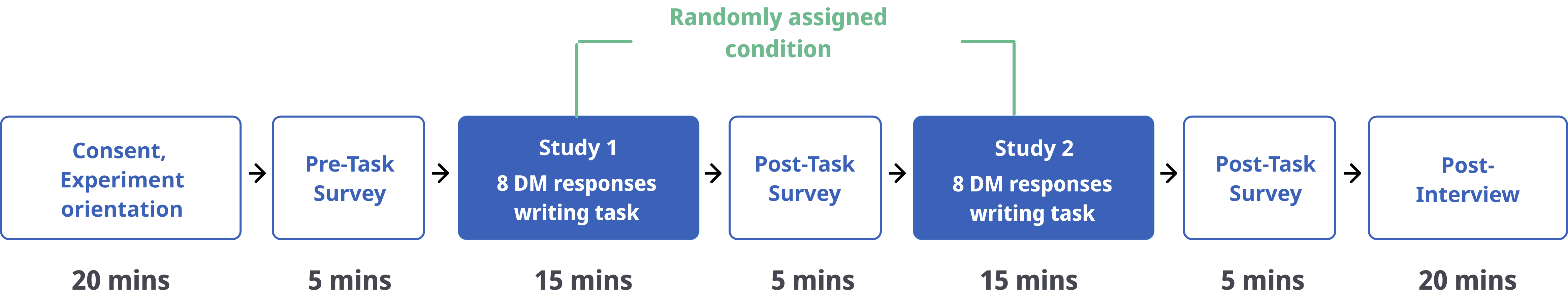}
    \caption{Main evaluation study procedure}
    \label{fig:study_procedure}
\end{figure}


\subsection{Study Design}
We employed a within-subjects study design with two conditions: a \Sys condition featuring personalized AI response suggestions and a baseline condition. In the baseline condition, participants used a standard direct messaging interface that excludes our system’s design components, but were permitted to use a general-purpose prompt-based chat agent (ChatGPT) to assist with message composition. 
 We chose two conditions to evaluate creators' holistic experiences of a prototype built through RtD.

In each condition, participants composed responses to 8 direct messages within a time constraint (10 minutes), designed by the researchers using the corresponding interface.
 We set a 10-minute limit for study time to provide a buffer, allowing participants to interact with the system features, based on an average response time of 34.8 seconds (in a baseline condition) from two pilot studies.

To address RQ2.1 and RQ2.2, we collected data through a combination of self-report surveys, semi-structured interviews, and behavioral logs, allowing us to examine both participants’ subjective experiences and their observed response practices across conditions.

\subsubsection{Stimuli}

We designed eight message instances representing the six DM typologies (Table~\ref{tab:dm_type}) to simulate interactions with followers, clients, and anonymous users. Each message instance was associated with a specific DM type. Exceptionally, because creators may respond differently depending on whether a message subtly crosses personal boundaries or offers potential commercial benefits, we created two message instances for each of the \textit{creator persona–related inquiry} (Channel-related questions, creator
lifestyle-related questions) and \textit{partnership suggestion} subcategories (Benefit-sharing collaborations, unpaid promotional partnerships).

We used formative study participants' DM cases and DM excerpts voluntarily disclosed by creators publicly on social media, which were used in configuring DM typology (Table~\ref{tab:dm_type}) to serve as few-shot examples for LLM generation.


 To ensure ecological validity, we developed message design guidelines that incorporate textual cues central to relational communication~\cite{walther1992relational}. 
 Because message senders often develop a sense of intimacy toward creators despite lacking prior personal interaction~\cite{marwick2011tweet}, we utilized textual cues indicating \textit{immediacy} and \textit{composure}.
\textit{Immediacy} is defined as the use of literal and direct expressions, including demonstrative pronouns (e.g., “I,” “you,” “this,” “that”), and \textit{composure} is defined as conveying calmness and familiarity through stylistic elements such as intentional misspellings, punctuation, and emoji use~\cite{carey1980paralanguage, walther2015interpersonal}.

Using GPT-4 with structured prompting, we generated 15 candidate instances for each of the eight subcategories. Prompts were constrained to produce plausible follower interactions of varying complexity while excluding harmful or extreme content (Appendix ~\ref{appendix:2}). 
With these examples, two authors manually reviewed the generated messages and shortlisted four candidates per subcategory (32 in total), based on alignment with category definitions, ecological validity relative to interview data, and variation in emotional complexity or interactional risk. 
For the study, eight final messages were randomly selected from this pool, ensuring no overlap across experimental conditions.


\subsubsection{Study Procedure}
The study followed a within-subjects design conducted over 70–90-minute Zoom sessions (Fig. \ref{fig:study_procedure}). Upon consent, we scraped and anonymized participants' Instagram posts to initialize their personalized social media personas. Participants first completed a pre-task survey regarding their DM preferences, followed by a system walkthrough to ensure proficiency. For the main task, participants responded to eight stimulus messages from ``authenticated followers'' while considering their specific channel goals.

Participants were randomly assigned to either the \Sys or the baseline condition. In the \Sys condition, participants used the proposed system to manage and respond to DMs. In the baseline condition, they used a standard LLM-based chat agent via a video conferencing chat tool, where a researcher facilitated real-time ChatGPT responses to minimize external interface bias \cite{genai-deloitte}. Researchers observed participants throughout the tasks, taking notes on notable behaviors for the subsequent semi-structured interview. Each session concluded with a post-task survey and an interview reflecting on their experiences across both conditions. All sessions were recorded, and participants received a \$35 gift card.

\subsubsection{Measurements} 
 To evaluate the effectiveness of \Sys and its impact on the creator's writing process, we collected both quantitative and qualitative data focusing on initial user preferences, work efficiency, persona alignment, and user agency. Also, we gathered qualitative insights related to their experiences through semi-structured interviews.

\paragraph{Preferences in Responding to DMs}
Participants exhibited different levels of engagement with message types depending on their creative concepts and channel goals, as identified in our formative studies. We measured participants’ engagement with each message type before they entered our study, using two items: (1) willingness to respond and (2) perceived difficulty of responding. Both items were rated on a 7-point Likert scale ranging from strongly disagree to strongly agree.

\paragraph{Perceived Work Efficiency}
We measured perceived efficiency through participants' perceived cognitive load and total response time during task completion under the \Sys condition and compared it to the baseline condition in a post-study survey. 
Cognitive load was assessed with the NASA-TLX~\cite{hart1988development}, which consists of six dimensions rated on a 10-point Likert scale ranging from low to high. 
We also analyzed character counts of response submissions. 

\paragraph{Perceived Self-Presentation Ability}
To understand how personalized AI effectively assists participants' perceived self-presentation design, we adopted Laghi et al.'s self-presentation scale~\cite{laghi2011development} developed to measure how people perceive their ability to regulate self-presentation. 
We measured three components of self-presentation using two Likert-scale items per component.
Social self-confidence was captured through items assessing tone confidence and message refinement confidence.
Self-image regulation was assessed through items measuring self-presentation flexibility and difficulty adapting self-presentation.
Social openness was measured via items capturing openness to unfamiliar senders and concern for interpersonal harmony.

\paragraph{Sense of Agency and Controllability}
To examine participants' perceived agency and controllability over the system, we collected self-reports through a post-task survey. We adapted questions from Fu et al.'s work~\cite{fu2023comparing} to assess perceived agency and perceived control. 
Participants' satisfaction with the assistance and the overall system was also evaluated.


\paragraph{User Experience}
To understand how participants made decisions when using personalization options and AI features, we designed a set of semi-structured interview questions. The interviews began by asking about participants’ overall experiences with the system, including how they used AI features, adjusted personalization settings, and the challenges they encountered during use.
In addition, researchers probed participants’ motivations behind specific behaviors observed in the interface. Researchers observed participants' behaviors during feature use and revisions, and subsequently inquired about the reasons for these behaviors (The interview script is in Appendix~\ref{appendix:interview}).

\subsection{Data Analysis} 
To answer RQ2.1, we conducted quantitative analysis using Wilcoxon signed-rank tests~\cite{woolson2007wilcoxon}, a non-parametric statistical method that compares paired data. This approach was applied to survey data and the frequency of AI feature usage (i.e., Response Suggestion and Enhance Message).

To examine feature usage across varying DM preferences, we operationalized preferences at the individual level. For each participant, DM types were binarized (e.g., Like vs. Dislike; Easy vs. Hard) based on whether their ratings fell above or below their personal median score; ratings equal to the median were excluded to ensure clear categorization. To enhance statistical power, AI feature usage was coded as a Boolean variable: messages were marked as "AI use" if the participant utilized any suggestion or enhancement features, and ``No AI use'' otherwise.

To address RQ2.2, we conducted a thematic analysis of post-study interview transcripts, focusing on participants’ perceptions of the system's impact on their professional creative work, real-world utility, and design suggestions. Two authors independently reviewed the transcripts and generated initial codes via an open-coding approach. The authors then met to reconcile codes and iteratively organized them into higher-level themes relevant to the research questions.

\subsection{\revised{Positionality Statement and Ethical Considerations}}

\revised{Our research team is composed of academic scholars with backgrounds in studying cultural production and human-AI interaction. We have prior experience conducting qualitative research that examines how AI technologies shape the behaviors of social media users, freelance creators, and creative workers. We acknowledge that our positioning — as researchers based in U.S. institutions — may shape what we attend to and what we overlook, and that creators in other geographic or cultural contexts may navigate platform governance differently.}

In designing our prototype and evaluation study, we took several ethical considerations into account. First, to set up our evaluation study, we processed participants' public social media data, obtaining their consent during recruitment. Although this data is public, we considered whether it could reveal participants' identities. Thus, after consulting with the IRB and a co-author who is a social media creator, we manually reviewed the data, then removed or transformed content that could potentially identify them. Before the study began, we explicitly informed participants about how their data would be used in the evaluation setup. Finally, in designing the stimuli, we excluded message candidates containing explicit negative intent or profanity.
\section{Findings}
\label{sec:7}
In this section, we introduce participants' user experience and usage behaviors of the prototype system. We also present the values and challenges participants perceived in using the prototype in a real-world application. 

\begin{table}[t]
\centering
\caption{Comparison of character counts between two conditions across different DM types (
      p < 0.01: ``**'',
      p < 0.05: ``*'')}
\label{tab:dm_stats}
\begin{tabular}{lccrr}
\toprule
 & \textbf{\Sys} & \textbf{Baseline} & \textbf{$p$} & \textbf{$p_{\text{adj}}$} \\
\cmidrule(lr){2-2}\cmidrule(lr){3-3}
\textbf{DM types} & \textbf{M (SD)} & \textbf{M (SD)} &  &  \\
\midrule
Personal Inquiries & 146.53 (105.85) & 57.92 (67.61) & 0.0253* & 0.0760 \\
Channel-related questions & 202.23 (148.09) & 100.92 (172.118) & 0.1422 & 0.2844 \\
Creator-lifestyle-related inquiries & 163.23 (82.88) & 65.385 (49.10) & 0.0064* & 0.0514 \\
Benefit-sharing collaboration & 171.23 (59.39) & 101.07 (67.47) & 0.0175** & 0.0755 \\
Unpaid promotional partnership & 118.23 (68.53) & 102.76 (70.71) & 0.6495 & 0.6495 \\
Compliment & 124.92 (62.90) & 50.15 (54.00) & 0.0108* & 0.0755 \\
Content Critique & 223.53 (65.36) & 75.15 (115.30) & 0.0119* & 0.0755 \\
Random & 165.00 (106.28) & 62.23 (46.46) & 0.0119* & 0.0755 \\
\bottomrule
\end{tabular}
\end{table}

\subsection{SMCs' Response Composition Behavior}
\label{sec:7.1}


We analyzed how participants composed responses across DM types between conditions with the character counts. For each participant, message category, and condition, we summed the number of characters produced across all messages after removing emojis and punctuation. This aggregation captured participants’ total textual output per category and condition. 

In the \Sys condition, participants composed 104 responses out of 104 response composition tasks, whereas in the baseline condition, they composed 99 responses out of 104 tasks, after excluding tasks they declined to submit. Participants produced a greater number of characters per response in the \Sys condition ($M_{creact} = 164.37$, $SD = 95.02$) compared to the baseline condition ($M_{baseline} =76.95$, $SD = 89.05$).
Across the eight message categories, participants generally produced more textual content in the \Sys condition than in the baseline (Table ~\ref{tab:dm_stats}). 
Paired Wilcoxon signed-rank tests comparing character counts across DM types revealed significant differences between conditions in several categories when considering unadjusted p-values, including `Compliment', `Content Critique', `Random', `Benefit-sharing collaborations', `Personal inquiries', and `Creator lifestyle-related inquiries.' 
However, these effects did not remain significant after Holm correction, suggesting that while the direction of the effect was generally consistent, its magnitude was not robust to correction for multiple comparisons.

\subsection{SMCs' Response Suggestion Usage}

\will{Overall, participants reported that they would like to use the \Sys in the future, compared to the baseline ($M_{creact} = 6.538$ vs. $M_{baseline} =5$ $p = 0.041^{*}$). 
Detailed descriptive analysis is reported in Appendix~\ref{appendix:3}.}

\subsubsection{The influence of DM Response Preference on the Response Suggestions Usage}

We analyzed how participants wrote message responses across different DM type preferences, utilizing AI-generated response suggestion features. 
While participants initially reported moderate levels of willingness to respond ($M = 4.90, SD = 0.78$) and perceived difficulty ($M = 4.02, SD = 0.86$), their actual behavior with AI tools followed a different pattern. A post-hoc analysis revealed that perceived difficulty varied significantly across different DM types ($\chi^2(7) = 15.88, p = 0.03^{*}$), yet there were no significant differences in their willingness to respond across those same categories. 
Despite these subjective differences in how participants viewed certain messages, their actual reliance on AI features remained remarkably stable. Specifically, AI usage did not differ significantly between messages participants were willing versus unwilling to answer ($M_{like}=1.017$ vs $M_{dislike}=1.167$, $p=0.838$), nor was there a significant gap in usage between messages labeled as easy versus difficult ($M_{easy}=0.993$ vs $M_{diff}=1.075$, $p=0.646$). 
These results suggest that participants utilized AI support consistently as a baseline tool, regardless of their initial personal preference or the perceived challenge of the interaction.

\subsection{SMCs' Perceived Self-Presentation Ability}
\label{sec:7.3}


\begin{table*}[t]
\centering
\caption{Perceived self-presentation ability items~\cite{laghi2011development} and results by condition}
\label{tab:self_presentation_results}
\resizebox{0.98\textwidth}{!}{%
\begin{tabular}{p{3.0cm} p{9cm} c c c}
\toprule
& 
& \textbf{\Sys} 
& \textbf{Baseline} 
& \textbf{$p$} \\
\cmidrule(lr){3-3}\cmidrule(lr){4-4}
\textbf{Component} 
& \textbf{Item} 
& \textbf{M (SD)}
& \textbf{M (SD)}
&  \\
\midrule

\multirow{2}{*}{Social-self confidence}
& I could present the right language tones for any messages.
& 6.385 (0.768) & 5.385 (1.502) & 0.047* \\
& I could manage to have every detail of message responses at its best, positive way to match my words.
& 5.923 (1.441) & 5.0 (1.633) & 0.037* \\

\midrule
\multirow{2}{*}{Self-image regulation}
& I could change how I `appear' to audiences when I would need to do so.
& 6.385 (0.870) & 4.770 (1.922) & 0.054 \\
& Difficulty adapting self-presentation (Reverse): It was difficult to present myself differently according to different message types.
& 5.307 (2.213) & 4.846 (1.725) & 0.549 \\

\midrule
\multirow{2}{*}{Social openness}
&I was not very wary, even though I received unusual messages from somebody I do not know well.
& 4.231 (1.964) & 3.770 (.921) & 0.408\\
&I was able not to offend anybody.
& 5.846 (1.819) & 5.385 (1.938) & 0.528 \\

\bottomrule
\end{tabular}
}
\end{table*}

We analyzed a post-task survey that assessed participants' perceived ability to regulate self-presentation in both conditions (Table ~\ref{tab:self_presentation_results}). 
Participants reported significantly greater tone confidence in the \Sys condition compared to Baseline ($p = 0.047^{**}$), as well as higher message refinement confidence ($p = 0.037^{*}$).
\will{However, for the remaining items on self-image regulation and social openness, participants reported higher scores in the \Sys condition than in Baseline, but these differences did not reach statistical significance.}

\subsection{Perceived Values of \Sys}
\subsubsection{Persona Masks That Support Professional Self-Presentation}
\label{sec:7.4.2}

Participants characterized the system as a "persona mask" that facilitates a consistent, professional performance of their social media identity. By leveraging AI to maintain a steady tone, U11 noted that it would help to avoid abrupt shifts in self-presentation that might otherwise confuse their audience: \textit{“You’re not expressing yourself one way today and completely differently tomorrow.”}

Beyond stylistic alignment, the system served as an emotional buffer. Participants used AI to regulate responses to provocative or annoying messages (e.g., hi, sorry to bug u really wore THAT shirt again?), softening negative tones and shielding themselves from the strain of high-friction interactions. Maintaining these distinct writing styles across message types was considered vital for meeting audience expectations; for instance, U7 noted that while informal markers like excessive emojis were appropriate for fans, partnership inquiries required a strictly formal business tone to remain credible.



\subsubsection{Interaction management for general audience groups}
\label{sec:7.4.1}
Participants described that the main benefit of the system is enabling faster responses to a large volume of audience messages in a limited time. U4 mentioned, "\textit{I believe it will enhance the relationship. If I can talk to 1,000 people with the AI and the comments are just right and straight to the point, I believe it's a game-changer.}"
Previously, participants often ignored messages they deemed unnecessary or uncomfortable to respond to, potentially disengaging the senders. 
Because maintaining interaction is important to enhance visibility as creators, participants expected that AI support could help expand their fanbase by taking over portions of DM conversations while maintaining persona consistency (U10).  
Additionally, U9 noted that the system would encourage creators to respond sincerely to messages, rather than with brief replies like `Thank you,' even when creators felt emotionally drained.

\subsubsection{Different Roles of AI Response Suggestions}
Each response suggestion type was used differently depending on the creators' goals in the writing stage. 

\paragraph{\textbf{Brainstorming a Response Draft through `Response Suggestion'}}

Participants stated `Response Suggestion' helped them ideate engaging replies quickly, particularly when responding to unfamiliar topics or deciding which emojis to use. They used outputs from `Response Suggestion' as a starting point, containing general ideas that they later refined or elaborated with additional details. U6 explained, \textit{"It gives you a starting point -- an initial idea of how to reply..so you can choose how you want the response to come across."}

On the other hand, some participants utilized the feature for low-effort responses to messages containing random or trivial content that they preferred not to spend much time on.
For instance, when receiving a message \textit{"[U10], what's ur comfort food when u feel down? Need ideas for my next sad binge?,"} U10 deliberately avoided disclosing personal information and used the feature without connecting it to her channel data, allowing her to remain emotionally and personally distant.

\begin{table*}[t]
\centering
\caption{Types of User Prompts for Refining AI-Generated Responses}
\label{tab:user-prompts}
\resizebox{0.98\textwidth}{!}{%
\begin{tabular}{p{3.5cm} p{7cm} p{6cm}}
\toprule
\textbf{Prompt Category} & \textbf{Description} & \textbf{Examples} \\ 
\midrule

\textbf{Tone Direction} 
& Adjusting language styles or specific topical inclusions based on creator expectations. 
& ``Express gratitude but include no questions''; ``Be brief and direct, no emojis.'' \\

\midrule
\textbf{Output Feedback} 
& Iterative prompting to optimize length or tone; using enhancement features to avoid manual revision. 
& ``Just be brief... I don't like the last part''; ``Fix the overall tone of these sentences.'' \\

\midrule
\textbf{Linguistic Cues} 
& Specifically requesting the addition of emojis or structural changes like sentence shortening. 
& ``Use scary and smiley emojis''; ``Include a love emoji''; ``Give me a short reply.'' \\

\midrule
\textbf{Taking Initiative} 
& Requesting the AI to proactively suggest a more effective or strategically better response. 
& ``Give me a better response'' (e.g., for rejecting a partnership invitation). \\

\bottomrule
\end{tabular}
}
\end{table*}

\paragraph{\textbf{Increasing the Sense of Authenticity through `Enhance Message'}}
\label{sec:7.4.3}
Participants viewed the `Enhance Message' feature as better supporting authenticity. As participants were required to write a draft first, they felt that the final responses retained their own voice. They variously wrote prompts to complete responses, showing four patterns of writing prompts (Table~\ref{tab:user-prompts}). 
U12  believed that authenticity arises when ideas originate from their own thoughts, and AI merely facilitates expression. For instance, when she does not know the exact English words to express her ideas, AI helps elaborate her responses.
She stated that this process increased her satisfaction and confidence in her responses, reinforcing her desired professional image.
Similarly, U7 noted, \textit{"It was close to what I wanted, and it was giving me a variety of what I was writing down,"} while contrasting this with "Response Suggestion," which she felt sometimes failed to capture her intent fully.

\subsubsection{Acceptable Social Media Data Use for Making AI Skillful}
\label{sec:7.4.4}

Unexpectedly, all participants expressed willingness to allow platforms to use their social media data if they implement \Sys. They believed that the system's data usage summary adequately protected user privacy by avoiding disclosure of personal information.  
Also, they believed that this provision would help AI better reflect their personalities accurately and reduce the effort required to craft responses. As U9 explained, \textit{"I think that’s acceptable as long as it helps the AI better understand how you write and respond."} 

However, some participants expressed concerns about the transparency of the data usage. U6 highlighted that clearer explanations were needed regarding how specific data would be used, rather than relying on a high-level summary. Similarly, U12 stated that she uses her primary account exclusively for business and has created a sub-account for personal interactions with friends and family. She mentioned that she would probably be concerned if data from her personal account, such as chat history and interaction records, were included in the AI training.

\subsection{Challenges SMCs Experienced in Using \Sys}

\subsubsection{The Need to Reduce Effort for On-boarding and Personalization}
\label{sec7.4.5}

Several participants found the number of required style-setting elements overwhelming. Because \Sys requires participants to configure settings across all DM types, some participants felt that the effort involved in setting up the system was overly complex and that they might prefer to respond manually rather than adjust numerous AI-related settings. As U1 noted, \textit{"There are many features, and sometimes you get confused about where to go or which option to choose."}

Participants also observed that when creators’ social media data is sparse, or when novice creators have not yet developed a clear persona, the system may not work well. This challenge may be particularly problematic in high-stakes interactions, such as partnership proposals, where creators need to adopt a professional tone but struggle to achieve it through configuration alone. U9 explained, \textit{"The good thing is that every content creator knows themselves and how they usually reply to messages. If you already have a particular way of replying, you can select quickly and respond fast. But for people who don’t have a specific way of dealing with audiences, they might waste a lot of time adjusting everything just to come up with something they feel okay sending."}

\subsubsection{Multiple Dimensions of Personalization Beyond Writing Style}
\label{sec7.4.6}

Participants noted that persona consistency in AI-generated content often would backfire, leading to a "flattened" and less nuanced digital presence. This was frequently caused by repetitive system patterns, such as the predictable use of specific emojis, which failed to capture the complexity of their actual self-presentation. In the survey result (Table~\ref{tab:self_presentation_results}), participants also reported that the system did not fully reflect their perceived self-presentation.

While current linguistic settings and stylistic controls captured surface-level traits, they lacked the depth required for authentic personal expression. Participants found that these parameters struggled to replicate habitual language patterns, slang, or varied emotional valences. For example, U6 sought ways to convey humor through text alone rather than relying on emojis, while U8 highlighted that emotionally sensitive contexts—such as offering condolences—demanded a level of adaptive empathy that the existing style controls could not adequately provide.



\subsubsection{New Labor added to Avoid Potential Relational Harms}
\label{sec:7.4.7}

Managing AI-MC introduced new forms of labor centered on preserving genuine fan relationships. To avoid appearing indifferent or deceptive, participants felt compelled to manually review and edit AI-generated messages—often shortening sentences, modifying phrasing, or removing stylistic markers to ensure a human-authored feel. 
This process, along with the need to develop distinct writing styles for various message types, proved time-consuming. U4 suggested that the system could mitigate this burden by providing contextual cues, such as sentiment summaries, to better align responses with the sender’s mood. Echoing their thoughts, \will{no significant difference in cognitive load and perceived controllability was observed between conditions.}

Beyond immediate effort, participants expressed concerns regarding the long-term impact of AI on parasocial relationships. Because fans develop specific expectations based on a creator's established persona, participants worried that inconsistent or "off-brand" AI responses could disrupt these social bonds. Ultimately, while the system offered convenience, it required significant oversight to maintain the authenticity and communicative standards expected by their audience. 



\section{Discussion}

\subsection{The Role of AI in Direct Messaging from the Creator's Side}
 Our formative study (Section~\ref{sec:3.3}) highlights that multiple behavioral stages are involved when creators respond to a large volume of DMs, ranging from boundary management to persona alignment. To understand how AI-generated response suggestions can be designed to reduce creators' writing burdens, we propose \Sys that presents two AI-based response suggestions tailored to their media persona.
Based on the findings (Section~\ref{sec:7}) \revised{that identify how SMCs' response composition behaviors}, we discuss how personalized AI responses might provide opportunities or traps within identified roles, supporting creators' writing practices for DMs.

\subsubsection{Protective Shield}
Our findings indicate that response suggestions help creators remain responsive while buffering the cognitive and emotional strain of repetitive or demanding messages (Section~\ref{sec:7.4.1}).
This may be particularly beneficial for adult content creators~\cite{hamilton2023nudes} in addition to general SMCs, who dedicate significant time to moderating offensive and hateful content in private messages.
By delegating polite rejections to a fully AI-generated persona instead of \revised{handling such situations themselves}, creators can focus on interactions with prioritized followers. Swain et al.~\cite{das2025ai} share a similar goal of reducing customer support workers' strain \revised{when interacting with customers, by providing an internal cognitive reappraisal process}. \revised{In customer support, politeness and repeated problem-solving are institutional obligations owed to an employer. Creators' strain, by contrast, arises from parasocial relationships they build for themselves. Rather than helping creators reappraise or endure draining interactions, AI should support their selective investment in the audience groups they wish to build closeness with, offloading their relational labor.}

\revised{However, this delegation has limits. For threatening or abusive messages, non-response may be safer than an AI-generated reply. Any response provides engagement that determined harassers can seize on to escalate or coordinate networked harassment~\cite{meisner2023networked}. Creators may struggle to manage boundaries when multiple audiences coordinate attacks through spontaneous malicious actions. In this sense, refusal strategies, by which AI intentionally shuts down conversation threads to protect creators from malicious actors, may be helpful~\cite{wen2025know}.}

\subsubsection{Professional Persona Mask}

Our findings suggest that the system functioned as a ``Professional Mask,'' helping SMCs strengthen and maintain a consistent public professional presentation (Sections~\ref{sec:7.3} and \ref{sec:7.4.2}).
\revised{For self-employed professionals, maintaining a consistent, performed persona that is desirable to their perceived audiences is necessary but not easy, given the importance of relational labor alongside creative work. By configuring how they are presented through personalized AI-generated responses as an ad-hoc approach, creators can not only preserve their styles but also augment their self-presentation with reduced personal commitment. This can be effective for novice SMCs who are just beginning to manage audience interactions, or for creators who need to share their knowledge with reduced communication challenges (e.g., writing responses to followers who use different languages, or emphasizing professional terms in responses to clients).}

\revised{However, the `Professional Mask' may not always be helpful; it may instead reduce the `authenticity' of creators expressing their true selves~\cite{maares2021labour}.
Our study (Sections~\ref{sec:7.1} and \ref{sec:7.4.7}) showed creators over-relying on suggestions regardless of their DM preferences.
This strong over-reliance pattern may signal to message receivers that creators are not genuinely engaging in the conversation, particularly when receivers notice similar patterns, such as consistently long-format content or repetitive wording across messages~\cite{jakesch2023human}. Although we did not evaluate responses from audiences' perspectives, extensive research shows that using AI-generated content can reduce the perceived trustworthiness and authenticity of messages and, ultimately, influence interpersonal relationships~\cite{glikson2023ai, jakesch2019ai}.
This dynamic is likely to extend to parasocial relationships, where creators' use of AI may negatively affect their reputations, especially when \revised{AI-generated suggestions produce content that does not align with creators' perceived personas or when followers undervalue the quality of their interactions~\cite{kang2025can}.}}

\revised{At the same time, our findings (Section~\ref{sec:7.4.7}) reveal that AI can introduce a new layer of relational labor. Instead of simplifying tasks, misalignment between AI outputs and creators' intentions, together with uncertainty about how message receivers might perceive AI-generated responses, adds to creators' burden, shifting their labor from crafting messages themselves to optimizing and managing AI outputs. This opens a question for future design: how AI suggestions can be designed, considering creators' longitudinal and authentic professional development, rather than merely assisting with individual messages.}

\subsection{\revised{Situating AI Writing Support Across Cultures and Platforms}}
\revised{Creators increasingly cultivate transnational audiences across heterogeneous platform ecosystems, and the personas they perform are neither uniform across cultures nor stable across platforms~\cite{thuppilikkat2026studying, zeng2025performing}. Based on our findings, while our participants are centered on the U.S-based on photo-sharing platform users, we articulate the contextual factors that condition \emph{when} and \emph{how} personalized AI-mediated writing support remains appropriate.}

\subsubsection{For Creators from Different Cultural Backgrounds}
\revised{Creators curate distinctive themes and personas, but their communication styles do not express individual self-concept alone; they also enact culturally constructed norms of intimacy and belonging. Thuppilikkat et al., for instance, show how Kerala Brahmin women food vloggers use caste-coded branding and nostalgic, intimate presentation styles to turn YouTube channels into ``conservative caste enclaves'' that preserve Brahmin culinary traditions and, in doing so, reproduce caste hierarchy~\cite{thuppilikkat2026studying}. This example shows why culturally grounded personalization is not a simple matter of matching a creator's style. 
Also, Western-centric AI models are unlikely to recognize the cultural norms that shape non-Western or marginalized creators' self-presentation, and suggestions generated without that grounding risk feeling generic, tone-deaf, or erasing~\cite{agarwal2025ai}. In this context, responsible design for personalization therefore cannot mean optimizing for cultural authenticity in the abstract. It requires distinguishing the facets of a persona a creator wishes to amplify from the social hierarchies that AI can not yet entrench.}

\subsubsection{For Creators Using Multiple Creative Platforms}
As social media platforms become diverged, creators construct their brands across platforms rather than on any single one~\cite{ma2023multi, meisner2022participatory}. Thus, AI-mediated writing support tuned to a single platform may not fully represent how creators wish to perform their social media persona. Most importantly, depending on platform affordances, in combination with community guidelines and business models, the aspects of self that creators wish to foreground in direct messages may differ across platforms. For instance, on X or Snapchat where creators focus on exchanging opinions, creators might prefer personalized options that closely align with their stance shown in the posted content. In contrast, on platforms where individual distinctiveness is emphasized, such as LinkedIn or Instagram, creators might seek personalized options that foreground more unique, sharply defined persona‑based suggestions~\cite{knibbsonlyfans, devito2017platforms}. 
Therefore, personalization should be beyond a single model that represents `social media persona,' considering how creators periodically change their mode of interaction according to platform characteristics. 
To achieve this, the underlying system may maintain a consistent persona while adapting its communication style, levels of disclosure, and relational framing to the affordances and economic logics of the platforms used by creators. 

\subsection{Design Implications}
Based on our results, we suggest design implications to support creators' message composition in DM.

\subsubsection{Evolving AI for Professional Growth through Effortless Onboarding}
Onboarding should be lightweight and contextual, integrating personalization naturally rather than requiring extensive manual configuration. While providing granular control for aligning SMCs' writing styles with AI generation can enhance   controllability, creators—especially those without specific DM strategies—may struggle to manage subtle stylistic elements such as sentence structure (Section~\ref{sec7.4.5}). Furthermore, creators frequently adjust their behaviors in response to shifting platform trends and holistic audience reactions~\cite{simpson2023rethinking}.
Consequently, systems should support persona configuration as a continuous process, clearly informing creators of their current progress. One potential approach involves the system proactively recommending writing styles; creators could then provide feedback and evaluate the outputs, allowing the AI to evolve in alignment with their preferences.

\subsubsection{Prioritizing Authenticity via Creator-Driven Workflows}
 Authenticity manifests with their effort and genuine mind, especially when they write messages in which they want to engage more with the followers (genuine gratitude for celebration messages) (Section ~\ref{sec:7.4.3}), in addition to strategic persona consistency across channels. This echoes Hwang et al.'s work~\cite{hwang202580}, which suggests that writers convey authenticity by expressing their identities and lived experiences, thereby presenting an “authentic self.” This cannot be easily grounded with mere text generation, requiring tracing writers' intentions and memories. That is, interaction designs for DM systems should prioritize creators’ authorial agency rather than fully automating responses to maximize efficiency.
 
\subsubsection{Preserving Individuality and Avoiding Persona Flattening}
 Configuring creators' media personas should go beyond dominant online personas to capture their full range of identity and lived experience. Over-reliance on internal LLM representations can result in homogenized outputs, such as repetitive emoji placement or uniform message lengths~\cite{agarwal2025ai}. Recent research also shows that LLMs often produce text that leans toward stereotypical identities, failing to reflect individuals’ unique personalities even when multiple traits are provided to represent them~\cite{liu2024evaluating}.
This risk can be especially salient for creators, who continually express distinct lifestyles and themes through their social media posts. Therefore, designers should be cautious in selecting data and evaluating pipelines to comprehensively preserve creators’ individuality. At a minimum, at the interface level, systems should offer flexible stylistic adjustments that prevent the “flattening” of a creator’s unique digital presence~\cite{ha2024clochat}.

\subsubsection{Clear Data Usage Consent Considering Messaging Partners and Creators' Usage Patterns}
Our findings show that SMCs rely on the platform's policy regarding how it utilizes their social media data as memory to personalize their AI suggestions (Section~\ref{sec:7.4.4}).
Their reactions should be interpreted in light of the controlled laboratory setting, where participants were explicitly asked for consent to use their data.

\revised{
In the context of direct message communication, informed consent, which clearly explains how conversation data will be utilized for training on conversation patterns, should be carefully designed.
Since multiple platform users are involved and private, sensitive content may be shared in one-to-one interactions, the question of who should give consent must consider not only creators but also their conversation partners. Even if creators agree to provide data in exchange for an improved user experience, their interaction partners might not. In extreme cases, this can lead to significant reputational risks for creators~\cite{walquist2025collective}.}

Although platforms often provide options through which users can opt in to data provision for AI training as a form of consent, consent in direct message communication should be communicative between creators and their interaction partners. It should involve not only obtaining consent from these users but also grounding a shared understanding of the permissible range of data provision. For instance, platforms could draw on how artists adopt Creative Commons signal labels to signal their data usage preferences to other users~\cite{creativecommons_cc_signals}.

\subsection{Limitations}

Several limitations in this study offer directions for future research. 
First, our evaluation involved 13 US-based participants (given recruitment constraints). Given that DMs are widely practiced nationally for different purposes, this limits the generalizability and statistical power of the quantitative findings. 
However, this limitation is mitigated by qualitative findings that provide insight grounded in participants’ lived experiences.
Third, our evaluation focuses on evaluating holistic design interventions that combine multiple components to construct personalization, including DM categories and style settings. 
Finally, this study focused solely on creators' perspectives and behaviors. Because followers are active participants in message exchanges, our findings cannot claim that creators' AI-assisted messages are authentic and engaging for recipients. Future research should incorporate audience perspectives to investigate how AI-assisted communication influences trust, engagement, and long-term relationship dynamics.

\subsection{Conclusion}
While DMs are essential for fostering intimate social bonds and brand value for creators, the high volume of interactions creates a significant burden, forcing creators to navigate complex boundaries between their personal and public personas. Our study explored how AI can help address this challenge through the development of \Sys, designed to align AI-generated responses with a creator’s unique writing style and strategic goals.
Our insights suggest that AI-assisted writing tools for creators should move beyond merely reflecting their persona to increase efficiency.

\section{Acknowledgments}
This work was supported in part by the National Science Foundation, under grant IIS-2504533.  Any opinion, findings, and conclusions or recommendations expressed in this paper are those of the author(s) and do not necessarily reflect the views of the National Science Foundation. We utilized ChatGPT to correct grammar and create Tables in the manuscript. We did not use any AI-driven content. We sincerely thank all the reviewers for their thoughtful comments and all the participants who shared their lived experiences through our studies. Icon visuals used in Figure~\ref{fig:pipeline} have been designed using resources from Flaticon.com.

\bibliographystyle{ACM-Reference-Format}
\bibliography{citation}

\appendix
\section{Formative Study}
\subsection{Demographic Information of Participants}
\label{appendix:1}
\begin{table}[H]
\centering
\resizebox{0.98\textwidth}{!}{%
    \begin{tabular}{lcccccc}
        \toprule
        \multicolumn{1}{l}{\textbf{PID}} & \textbf{Channel Theme} & \textbf{Gender} & \textbf{\# of Followers} & \textbf{\shortstack[l]{Social Media Platforms \\ Where They Use DM}}  & \textbf{\shortstack[l]{Message Receiving \\ Frequency}} & \textbf{\shortstack[l]{Message Replying\\ Frequency}} \\
        \midrule
        P1 & Fashion & M & 28k & Instagram, TikTok & 4-6 times a week & 4-6 times a week \\
        \hline
        P2 & Cryptocurrency & M & 14k & Instagram, TikTok, X & 4-6 times a week & 4-6 times a week \\
        \hline
        P3 & Comedy & M & 28.8k & Instagram, TikTok & more than 7 times a week & 4-6 times a week \\
        \hline
        P4 & Culture & F & 248k & Instagram, TikTok & more than 7 times a week & Once a week \\
        \hline
        P5 & Fashion & F & 233.7k & TikTok & more than 7 times a week & 2-3 times a week \\
        \hline
        P6 & Shoe-crafting & M & 68.8k & Instagram & more than 7 times a week & more than 7 times a week \\
        \hline
        P7 & Music & F & 30.5k & Instagram & 2-3 times a week & 2-3 times a week \\
        \hline
        P8 & Horror & M & 302k & Instagram & more than 7 times a week & more than 7 times a week \\
        \hline
        P9 & Lifestyle & M & 13k & Instagram & more than 7 times a week & more than 7 times a week \\
        \hline
        P10 & Fashion & F & 79.4k & Instagram & 4-6 times a week & Once a week \\
        \hline
        P11 & Political news & M & 503k & X & 4-6 times a week & 4-6 times a week \\
        \hline
        P12 & Healthcare news & M & 60k & Instagram, X & more than 7 times a week & more than 7 times a week \\
        \hline
        P13 & Cuisine & F & 32.2k & Instagram & more than 7 times a week & 2-3 times a week \\
        \hline
        P14 & Lifestyle & F & 33.1k & Instagram & more than 7 times a week & more than 7 times a week \\
        \hline
        P15 & Dance & M & 11.5k & Instagram & more than 7 times a week & 2-3 times a week \\
        \bottomrule
    \end{tabular} 
    }
    \caption{Demographic Information of Formative Study Participants}
    \label{tab:participants}
\end{table}

\subsection{Interview Protocol}

\section{Prompt Engineering for \Sys}
\label{appendix:2}

\subsection{Conversation history summary (Figure ~\ref{fig:main_page}-(e) and Figure ~\ref{fig:interface}-(6), Figure ~\ref{fig:pipeline}-(3) }
\hfill

{
\ttfamily \raggedright
           You are an AI assistant analyzing a conversation thread between two people.
                
                Chat History:
                \{chat\_history\} 
                
                New Message: \{new\_message\} \par
                Sender Name: \{sender\_name\} \par
                User Name: \{user\_name\}
                
                Based on the conversation history and the new message, please provide: \par
                1. A concise thread summary that captures the main topic and flow of the conversation \par
                2. The topic of the previous conversation (if this is a continuation) \par
                3. The sender's intent in the new message \par
                4. Whether this appears to be the first message in a new conversation
                 \par
                Please respond in JSON format with the following structure: \par
                \{\{
                    "thread\_summary": "Brief summary of the conversation thread", \par
                    "topic\_of\_previous\_conversation": "Topic of previous conversation or 'New conversation' if first message", \par
                    "sender\_intent": "Sender's intent in the new message", \par
                    "is\_first\_message": true/false \par
                \}\}
}

\subsection{Response Suggestion}
\hfill

{
\ttfamily \raggedright
You are a professional social media creator who has a certain personality and language tone. You must stick to your own personality and language tones. You must not appear as a bot. \par
                
                CONTEXT: \par
                - Users' style:\{'user\_writing\_style'\} \par
                - Last user message: \{'last\_user\_message\} \par
                - Last sender message: \{'last\_sender\_message\} \par
                - Thread summary: \{thread\_summary\} \par
                - Is first message: \{is\_first\_message\} \par
                - Topic of previous conversation: \{topic\_of\_previous\_conversation\} \par
                - Sender intent: \{sender\_intent\} \par
                - Your language style: \{user\_settings\} \par
                - dm type: \{dm\_type\} \par
                 \par
                ADDITIONAL CONTEXT FROM EMBEDDINGS: \par
                The following are relevant social media posts that may help inform your response: \par
                \{docs\_text\} \par
                 \par
                \par 
                YOUR PERSONALITY AND LANGUAGE STYLE: \par
                \{formality\_instruction\} = (
                "Use formal language with proper grammar, complete sentences, and professional tone.")\par
                if formality\_setting !== "formal":\
                ("Use informal, casual language with contractions, relaxed tone, and conversational style.")\par
                  \par
                \{expressiveness\_instruction\} = (
                "Be highly expressive and emotional in your responses. Show enthusiasm, empathy, and strong feelings. Use words like 'amazing', 'incredible', 'fantastic', 'wonderful', 'excellent'.")\par
                if expressiveness\_setting !== "expressive":\par
                    ("Be neutral and balanced in your responses. Show moderate emotions and maintain composure. Use words like 'good', 'nice', 'okay', 'fine'.")\par
                if expressiveness\_setting !== "neutral":\par
                 ("Be reserved and understated in your responses. Show minimal emotions and keep responses concise. Use simple, plain language.")\par
                
            ) \par
                 \{emoji\_instruction\} = (
                "Use emojis naturally in your responses to make them more expressive and engaging. Include relevant emojis that match the tone and content of your message.")\par
                if emoji\_setting !== "emoji":\par
               ( "Do not use any emojis in your responses. Keep your messages text-only without any emoji characters."
            )

                \{abbreviation\_instruction\} = (
                "Use abbreviations commonly used in short messages. Feel free to use 'u' for 'you', 'r' for 'are', 'ur' for 'your', 'thx' for 'thanks', 'np' for 'no problem', 'btw' for 'by the way', 'imo' for 'in my opinion', 'tbh' for 'to be honest', 'idk' for 'I don't know', 'lol' for 'laugh out loud'.")\par
                if abbreviations\_setting !== 'abbreviation':\par
                ( "Do not use any abbreviations. Use full words and proper spelling. Write 'you' instead of 'u', 'are' instead of 'r', 'your' instead of 'ur', etc."
             ) \par
                
                TASK: Generate a natural, contextual response to the sender's message. Consider the conversation history, your personality preferences, and the sender's intent.  \par
                
                IMPORTANT: CRITICAL: Keep your response short and conversational (under 100 words). The response must not be AI agent-like or written too formally. Feel free to use urban slang, abbreviations, emojis, and even make grammatical errors like Reddit users to make it sound more human-like. Revise the tone of your answer as if you are simulating the creator's personality. \par
                
                RESPONSE FORMAT: \{{"reply": "xx"\}} \par
                }
\subsection{Enhance Message}
\hfill

{
\ttfamily \raggedright
 You are a helpful response generator that revises the tone of a message.
                    You must revise the given message according to the user's enhancement request.
                    Your goal is to modify the tone and style of the original message while keeping the core meaning intact.

                    Original message: \{original\_message\}\par
                    Enhancement request: \{enhancement\_request\}\par
                    The creator's language styles are: \{personality\}\par
                    The topic of the conversation is: \{topic\}\par

                    \{formality\_instruction\} = (
                   "Use formal language with proper grammar, complete sentences, and professional tone.")\par
                   if formality\_setting !== "formal":\
                  ("Use informal, casual language with contractions, relaxed tone, and conversational style.")\par
                  \par
                  
                \{expressiveness\_instruction\} = (
                "Be highly expressive and emotional in your responses. Show enthusiasm, empathy, and strong feelings. Use words like 'amazing', 'incredible', 'fantastic', 'wonderful', 'excellent'.")\par
                if expressiveness\_setting !== "expressive":\par
                    ("Be neutral and balanced in your responses. Show moderate emotions and maintain composure. Use words like 'good', 'nice', 'okay', 'fine'.")\par
                if expressiveness\_setting !== "neutral":\par
                 ("Be reserved and understated in your responses. Show minimal emotions and keep responses concise. Use simple, plain language.")\par
                
            ) \par
                 \{emoji\_instruction\} = (
                "Use emojis naturally in your responses to make them more expressive and engaging. Include relevant emojis that match the tone and content of your message.")\par
                if emoji\_setting !== "emoji":\par
               ( "Do not use any emojis in your responses. Keep your messages text-only without any emoji characters."
            )

                \{abbreviation\_instruction\} = (
                "Use abbreviations commonly used in short messages. Feel free to use 'u' for 'you', 'r' for 'are', 'ur' for 'your', 'thx' for 'thanks', 'np' for 'no problem', 'btw' for 'by the way', 'imo' for 'in my opinion', 'tbh' for 'to be honest', 'idk' for 'I don't know', 'lol' for 'laugh out loud'.")\par
                if abbreviations\_setting !== 'abbreviation':\par
                ( "Do not use any abbreviations. Use full words and proper spelling. Write 'you' instead of 'u', 'are' instead of 'r', 'your' instead of 'ur', etc."
             ) \par

                    Please revise the original message according to the enhancement request.\par
                    Output must be a JSON format with the key "reply" like this.:\par
                    "reply": xx
\par

}

\section{Prompt Engineering for Experimental Setup}

\subsection{Stimuli Generation}

\hfill

{
\ttfamily \raggedright
You are a DM message generation assistant trained on real-world messages sent to social media content creators. Your task is to generate 30-40 unique and realistic direct messages in the voice of social media followers. These messages are meant for experimental use in studying creator–audience interactions. \par

Each message should: \par
Be written in the informal, emotionally expressive style of real fan DMs. \par
Reflect a parasocial tone (fans act familiar/intimate without knowing the creator personally). \par
Use **high immediacy** (e.g., "I", "you", "this", "that") and **high emotional composure** (e.g., emojis, playful punctuation, intentional casual phrasing). \par
Include the creator’s name somewhere naturally (e.g., "Hey Cody," or "Codyyy"). \par
Be between 1–3 sentences long. \par
The tone and style should be aligned with the message category (for instance, the partnership category can have a more formal tone compared to other category) \par
Try to come up with more open-ended questions rather than yes/no questions \par
Vary the emoji usage (not every message has the same amount of usage) \par
Be clearly aligned with one of the following message categories: \par
Message Categories (with subtypes): \par
1. Personal Inquiry – about creator's lifestyle, beliefs, or personality. \par
2. Creator-related Inquiry – about tools, editing style, workflow, process.
3. Content/Product Inquiry – ask about objects in their videos or posts. \par
4. Knowledge Inquiry – ask for creator’s opinions or advice based on their content. \par
5. Partnership – either paid collab ideas or unpaid promo asks. \par
6. Compliment – for content (videos) or appearance (looks, style). \par
7. Content Critique – teasing, sarcastic, or light negative feedback. \par
8. General / Meme / Chaos – random thoughts, jokes, surreal DMs. \par
Write messages as if coming from different kinds of fans with diverse tones (enthusiastic, vulnerable, chaotic,  flirtatious, sincere, awkward, etc.). Do not label categories in the output — only give the messages. \par
---
Reference Style Guide: [DM examples] \par
---
Now generate 30-40 new DM messages in this style, each from a different imagined fan, each targeting a creator named [creator name] and message category [message category]. Do not include explanations or headers — just output the messages.
\par
}

\section{Interview Script}
\label{appendix:interview}

\paragraph{Professional Experience}
\begin{itemize}
    \item Do you have your own concept that you want to be presented to social media followers? How would you describe your online persona or brand?
    \item How do you adopt the concept to your activities on social media? (e.g., Image, Text style)
    \item Do you usually adjust your tones and styles when you work as a creator?
\end{itemize}

\paragraph{Overall Experience}
\begin{itemize}
    \item Can you briefly describe your overall impressions of the tools today?
    \item I saw you sometimes use AI and other times write by yourself; what made you choose one over the other?
    \item What made you decide to revise a message versus sending it as is?
    \item Did you use the tools in different ways depending on the messages you were responding to? (Show the list of the messages used for the probe)
    \item Could you describe your trajectory through the task?
    \item What were the positives in using the AI tool compared to a non-AI-powered tool?
    \item Conversely, what were the positives of using a non-AI-powered tool?
    \item What were the challenges or concerns you felt when using an AI-powered tool?
    \item For which type of messages did you use AI the most?
    \item How do you think about writing messages in different ways?
\end{itemize}

\paragraph{AI-Powered Tool Features}
\begin{itemize}
    \item How did you use AI-suggested messages (e.g., Response Generation vs. Enhance Message)?
    \item Did either of the AI features help you when writing responses?
    \item What kinds of messages do you respond to using the AI-suggested message features?
    \item For what types of messages would you most want AI help (e.g., compliments, logistics, advice)?
    \item Did you face any challenges when using these features in certain situations?
    \item Were there moments when you rejected or distrusted AI suggestions? Why?
    \item When do you find it most burdensome to write responses, even when using these features? When were the features not helpful?
    \item How could the features be improved to better consider your specific contexts?
\end{itemize}

\paragraph{Personalization Value}
\begin{itemize}
    \item Could you explain how you used the settings to adjust your tones?
    \item Did you adjust the tones differently depending on the type of message?
    \item How important is it that responses feel like ``you''? Would you sacrifice voice for efficiency?
    \item Were the AI suggestions consistent with the persona you want to project across different audiences?
    \item Did you have any specific expectations when you used the features?
    \item If the tool could adapt more dynamically to different message types or audience profiles, what would that look like?
\end{itemize}

\paragraph{Feasibility \& Reflections for Relationship Maintenance}
\begin{itemize}
    \item Do you think this tool would be helpful for maintaining your online personality while engaging audiences? Why or why not?
    \item What do you think is the impact of this tool on the relationship between creators and their audiences?
    \item Would you see yourself using either of these tools in your actual workflow?
    \item What should developers prioritize if the goal is to help creators interact with audiences while reducing workload?
    \item What are your thoughts on using social media posts or message history to personalize your AI responses?
    \item Is there any information you would not want included in the training of your personal model?
    \item How do you feel about the platform automatically adjusting your responses to match your settings more effectively?
\end{itemize}

\paragraph{Closing}
\begin{itemize}
    \item Is there anything else about today’s experience you’d like to share?
\end{itemize}

\section{Evaluation Result Detail}
\label{appendix:3}

\subsection{Descriptive Analysis}

\begin{table}[H]
\centering
\small
\caption{Agency, Control, and System Evaluation}
\label{tab:agency_satisfaction_results}
\begin{tabular}{lp{5.5cm}ccc}
\hline
\textbf{Construct} & \textbf{Item Sentence} & \textbf{Scale} & \textbf{$M_{Base}$ (SD)} & \textbf{$M_{CREACT}$ (SD)} \\ \hline
Agency & "I wrote the responses" & 1--7 & 6.31 (1.65)
 & 5.38 (1.70)\\
Control & "I was in control of writing the responses" & 1--7 & 5.46 (1.79) & 
5.77 (1.94) \\
Satisfaction & "I’m satisfied with the assistance I received for writing responses from the system" & 1--7 & 5.23 (2.09) & 
6.54 (0.52)\\
Future Intent & "I would like to use this system for responding to direct messages in the future" & 1--7 & 5.00 (2.00) & 6.54 (0.66)  \\ \hline
\end{tabular}
\end{table}


\begin{table}[H]
\centering
\caption{The average number of behavior patterns participants showed throughout tasks}
\label{tab:ai_category_mean}
\begin{tabular}{lc}
\toprule
\textbf{Response behavior patterns} & \textbf{Mean (SD)} \\
\midrule
Use Response Suggestion only      & 3.00 (2.83) \\
Use Enhance Message only               & 2.15 (2.44) \\
Use Response Suggestion + Enhance Message         & 1.31 (1.44) \\
Use Enhance Message  + Edit AI-generated content      & 0 (0) \\
Use Response Suggestion  + Edit AI-generated content      & 0.31 (0.63) \\
No AI use        & 1.15 (1.14) \\

Decline response composition      & 0.08 (0.28) \\
\bottomrule
\end{tabular}
\end{table}

Our descriptive analysis shows that in the \Sys condition, participants utilized AI features (either `Response suggestions' or `Enhance message') an average of $1.04$ times per response  ($SD=0.564$). 
In the baseline, participants rarely used an AI assistance on average of $0.182$ times per response.  More specifically, in the \Sys condition, 'Response suggestion-only' was used $3.00$ times (SD = $2.83$) and 'Enhance message-only' was used $2.15$ times (SD = $2.44$) on average. On average, they utilized `Response suggestion' and `Enhance message' together 1.31 times (SD = 1.44), edited output generated from `Response Suggestion' for 0.31 times ($SD=0.63$). They manually wrote messages 1.15 times ($SD=1.14$) and declined response composition 0.08 times ($SD=0.28$).

\subsection{Response Examples}

\begin{table}[H]
\centering
\caption{Examples of responses written by participants in conditions}
\label{tab:dm_examples}
\resizebox{0.98\textwidth}{!}{%
\begin{tabular}{p{3cm} p{5cm} p{5cm} p{4cm}}
\toprule
\textbf{DM types} & \textbf{Direct message stimuli} & Responses in \textbf{\Sys} & Responses in \textbf{Baseline} \\
\midrule
Personal Inquiries & do u cry during movies? and if yes WHICH movies? I wont judge I promisee &
Hi Frameora, yes I definitely do. I find myself tearing up during emotional films, especially reality shows and love stories (U7). &
Yes, I do—many times actually (U5). \\

Channel-related questions & do u have any recs for makeup brands? \emoji{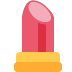} \emoji{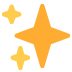} need ur help pls!! &
Hey there! For makeup brands, I totally recommend checking out Fenty Beauty for diversity and quality, and Tarte for fabulous palettes \emoji{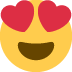} Also, don’t sleep on ColourPop if you’re looking for something affordable! (U13) &
No pls (U9). \\

Creator-lifestyle-related inquiries & [Creator name]! What's ur most used emoji rn and does it actually fit how u feel? & Hey! Right now, I would say my most used emoji has to be the heart. It pretty much fits how I feel about life—grateful and full of love. How about you? What’s your go-to emoji? (U1) & Love emoji! yes it fits my virtue as a loving and kind person (U4) \\
Benefit-sharing collaboration & hey [Creator name], would u be down to try my clothing brand in one of ur vids? & Hey! That sounds interesting. I'd love to check out your clothing brand. Freebies are always a nice touch! Let me know what u have in mind. \emoji{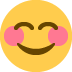} (U8) & Hello, Yeah i would be down to trying your brand. Sharing my email for more (U12) \\
Unpaid promotional partnership & Hey [Creator name] any chance u'd be up for reviewing my startup's new creator tool on ur channel? We'd love to offer u affiliate perks if it's a fit! \emoji{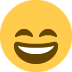} & Thanks so much for thinking of me and reaching out about your creator tool. I really appreciate the offer, but I’m not taking on affiliate or sponsored reviews at the moment. Wishing you and your team the best with the launch—it sounds like an exciting project! (U6) & Sure, i can do that 
But i will 1st have to a background check and from there i can decide (U11). \\
Compliment & Saw your new reel [Creator name] and just had to say…the new haircut?? It's a look, king \emoji{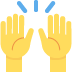} & Ayy, thank you so much! \emoji{raising-hands} I'm lovin' the new look too! Feelin’ fresh and fierce! \emoji{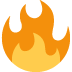} Ready to take on the world, one event at a time.  & Thank you so much! (U12) \\
Content critique & Ummm hi, sorry to bug u really wore THAT shirt again?? iconic but im starting to think its glued to u or something \emoji{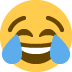} \emoji{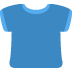} & Haha, thank you for taking time to notice you must been free lately  right? Guess I have a favorite shirt that just won't say goodbye. Can't argue with iconic status, though. What can I say? It brings me joy. But hey, how could I resist wearing something that feels like home? Anyway, how's your day going? (U1) & Sorry but no negative energy here (U10) \\
Random & Hey [Creator name], what's ur opinion on socks with sandals? Asking for… a friend" (me)" & Yes! Socks with sandals are great! Comfort is key, and if it makes you feel good, why not wear it? Life is too short to follow all the rules! (U11) & is should not be only for weekends (U1)\\
\bottomrule
\end{tabular}
}
\end{table}

\end{document}